\documentclass{aa}

\usepackage{graphicx}
\usepackage{txfonts}
\usepackage{lipsum}
\usepackage{subcaption}
\usepackage{lscape}             
\usepackage{placeins}           

\usepackage{hyperref}

\usepackage{natbib}

\usepackage{multicol}
\usepackage[dvipsnames]{xcolor}

\def\msun{\ifmmode M_{\odot} \else M$_{\odot}$\fi}
\def\msunyr{\ifmmode M_{\odot} {\rm yr}^{-1} \else M$_{\odot}$ yr$^{-1}$\fi}
\def\zsun{\ifmmode Z_{\odot} \else Z$_{\odot}$\fi}
\def\lsun{\ifmmode L_{\odot} \else L$_{\odot}$\fi}

\newcommand{\ha}{\ifmmode {\rm H}\alpha \else H$\alpha$\fi}
\newcommand{\hb}{\ifmmode {\rm H}\beta \else H$\beta$\fi}
\newcommand{\hg}{\ifmmode {\rm H}\gamma \else H$\gamma$\fi}
\newcommand{\hd}{\ifmmode {\rm H}\delta \else H$\delta$\fi}
\newcommand{\mstar}{\ifmmode M_\star \else $M_\star$\fi}
\newcommand{\teff}{\ifmmode T_{\rm eff} \else $T_{\rm eff}$\fi}

\def\Oii{[O~{\sc ii}] $\lambda$3727}
\def\oiii{[O~{\sc iii}]}

\def\Hei{[He~{\sc i}]$\lambda$4027}

\def\Oiiill{[O~{\sc iii}] $\lambda$4364}
\def\Heii{[He~{\sc ii}]$\lambda$4687}
\def\Neiii{[Ne~{\sc iii}]$\lambda\lambda$3969, 3870}

\begin{document}

   \title{Tracing the Evolution of the Balmer Break from Cosmic Dawn to Cosmic Noon with JWST}

   \subtitle{}

   \author{Adarsh Kuruvanthodi
          \inst{1,2}\thanks{Corresponding author: adarsh.kuruvanthodi@unige.ch},
          Daniel Schaerer \inst{1,3},
          Rui Marques-Chaves \inst{1},
          Andrea Weibel \inst{1},
          Damien Korber \inst{1},
          \and
          Corinne Charbonnel \inst{1,3},
          }

   \institute{Department of Astronomy, University of Geneva, Chemin Pegasi 51, 1290 Versoix, Switzerland
         \and 
                 Department of Physics, Indian Institute of Science Education and Research (IISER) Tirupati, Yerpedu, Tirupati - 517619, Andhra Pradesh, India
         \and
                 CNRS, IRAP, 14 Avenue E. Belin, 31400 Toulouse, France}

\authorrunning{Kuruvanthodi et al.}

   \date{}

  \abstract
   {The Balmer break (BB) is a key spectral feature for constraining stellar population ages, star formation histories, and redshifts of high-redshift sources. The redshift evolution and distribution of BB strength, together with the properties of BB galaxies, constrain stellar population characteristics and the nature of star formation across cosmic epochs. However, a systematic and unbiased characterization of BB strengths across the full galaxy population remains limited with the James Webb Space Telescope (JWST).}  
   {We aim to characterize the redshift evolution of BB strength over $z=3.5$–10 and its distribution across different epochs using photometry. We also examine correlations between BB strength and key physical parameters within these redshift intervals. We further assess the implications of BB galaxies for the nature of star formation and stellar populations at $z>3.5$.}
   {We used the JWST NIRCam photometric observations taken as part of various programs, including CEERS, JADES, FRESCO, and PRIMER. We estimated the BB strength of the objects with two adjacent broadband filters in various redshift windows between redshifts 3 and 10, which exclude strong line contamination. We employ the SED-fitting code CIGALE for both SED fitting and the generation of mock galaxy simulations.} 
   {We find that the median Balmer break strength (expressed as a flux ratio) increases from cosmic dawn to cosmic noon, from 1.1 to 1.5, primarily driven by the age of the stellar population. These estimations are in agreement with the latest spectroscopic estimations in the literature. We identify objects with extremely large BB strengths (BB$>3.0$) at $z=3.5$–4 and $z=7$–10, indicating strong extinction combined with an old stellar population and the presence of Little Red Dots (LRDs) in the former, and predominantly LRDs in the latter. Considering a constant star formation (instantaneous burst) history, we find that the average age of the stellar population decreases from 350 (50) to 20 (10) Myr at $z=3.5-10$. We find significant positive correlations between BB strength and stellar mass, age, UV slope ($\beta$), and extinction, and negative correlations with $EW(\mathrm{H}\alpha)$ and $EW(\mathrm{H}\beta)$ across redshifts. A negative correlation between BB strength and sSFR is observed in certain redshift intervals; however, no significant correlation is found with SFR, absolute V-band magnitude ($M_{V}$), or absolute UV magnitude ($M_{\mathrm{UV}}$). Overall, we highlight the effectiveness of BB as a standalone diagnostic of stellar populations and star formation in the early universe.}
   {}
   \keywords{Galaxies: evolution -- Galaxies: formation -- Galaxies: general -- Galaxies: high-redshift -- Galaxies: star formation -- dark ages, reionization, first stars}

   \maketitle
   \nolinenumbers

\section{Introduction}

Galaxies with little or no star formation have been of great interest in the era of the James Webb Space Telescope (JWST). A significant number of such quenched or temporarily quenched (dormant) galaxies have been spectroscopically identified recently, even above redshift $z > 7$ \citep[e.g.][]{2024Natur.629...53L, 2024MNRAS.529.1299V,  2024A&A...691A.310K, 2025ApJ...983...11W}. For all these sources, the quiescent nature is revealed by the presence of a Balmer break (BB). Interestingly, such galaxies have been observed in all mass ranges, unlike the local universe, where such galaxies are prominent in the massive (>$10^{10}\msun$) galaxy regime \citep{2024Natur.629...53L}. This highlights the importance of studying Balmer break galaxies, regardless of their mass, in order to understand the nature of star formation in the early Universe.

The Balmer break and D4000 break around 4000\AA\ are key spectral feature for efficiently identifying the quiescent population of galaxies \citep{1999ApJ...527...54B, 2003MNRAS.341...33K, 2020ApJ...897...44S}. In the pre-JWST era, however, these features have been spectroscopically identified only up to redshift 3.7 \citep{2017Natur.544...71G}. Historically, this feature has extensively been used to constrain the age and star-formation histories of quiescent galaxies \citep{1999ApJ...527...54B, 2003MNRAS.341...33K, 2019ApJ...874...17B, 2019MNRAS.490..417C}. On the other hand, when galaxies are in an actively star-forming phase, their spectra lack a Balmer break and often show a Balmer Jump instead, due to nebular emission. The nature and strength of the Balmer feature is therefore an important tool to understand the overall star-forming nature of galaxies and their history. Both \citet{Langeroodi+2024a} and \citet{2024ApJ...976..193R} showed that an average stellar population exhibits a Balmer break feature up to a redshift of 5–6. Beyond this, the feature moves to the Balmer jump. This indicates, on average, the galaxies are in a star-forming phase above $z\sim6$, while the quiescent population becomes significant below $z\sim5-6$.

In the past, the Balmer break has often been associated with massive quiescent galaxies. Recently, the BB has been observed in a wide range of objects, including quiescent galaxies, post-starburst galaxies \citep[e.g.][]{2019ApJ...874...17B, 2023Natur.619..716C, Glazebrook+2024a, 2025NatAs...9..280D}, mini-quenched (temporally quenched) galaxies \citep[e.g.][]{2024Natur.629...53L}, dusty galaxies \citep[e.g.][]{2024ApJ...974..145S}, and newly discovered objects like Little Red Dots \citep[LRDs;][]{2024ApJ...963..129M, 2025ApJ...978...92L}. LRDs are observed up to a redshift of 10, and so far, the strongest BB, exceeding the maximum predicted by normal stellar populations, has been reported in an LRD at $z \sim 7.8$ \citep{2025arXiv250316596N}. The nature of LRDs and the origin of the BB in these sources is still debated. Possible explanations for strong BBs include dense gas around an active galactic nucleus \citep[AGN;][]{2025ApJ...980L..27I, 2025MNRAS.tmp.1770J, 2025arXiv250316596N}, a black hole star \citep[BH$\star$;][]{2025arXiv250316596N}, or supermassive stars \citep[SMS;][]{2025A&A...701A.168D, 2025arXiv250712618N,Chisholm2026Little-Red-Dots}. 
\looseness=-1

Classical quiescent galaxies are often classified into two categories depending on their quenching pathways. Some of them undergo a sudden decrease in their star formation due to the rapid ($\sim 100$ Myr) depletion of available gas. Such a fast quenching can often be associated with a merger-induced starburst or quasar-mode AGN activity, and the resulting galaxy evolves into a post-starburst phase showing Balmer break and Balmer absorption features in its spectrum \citep[e.g.][]{2019ApJ...874...17B}. On the other hand, in the slow quenching case, the galaxy depletes its gas reservoir in a slow phase lasting up to 1-4 Gyrs. Such a situation arises when the dark matter halo of the galaxy exceeds $10^{12}\msun$, as the infalling gas will be shock-heated, which makes cooling inefficient and star formation difficult \citep{2003MNRAS.345..349B, 2005MNRAS.363....2K}. Low accretion (radio mode) AGN feedback and environmental effects can also lead to or assist in slow quenching \citep{2015Natur.521..192P}. In slow quenching, the galaxies are expected to show a D4000 break after the quenching phase. This is not a continuum feature but rather the shape of the SED, and the presence of various absorption lines creates a break-like feature around 4000\AA\ . Number density studies often suggest the slow mode of quenching becomes the dominant quenching pathway at low redshift compared to fast quenching and vice versa \citep{2019ApJ...874...17B}.

On the other hand, a relatively weak Balmer break has been observed in a significant proportion of low-mass galaxies (<$10^9\msun$) at high redshift. This scenario became particularly interesting when \citet{2024Natur.629...53L} presented a Balmer break galaxy in the low mass regime with a spectrum showing no emission lines and a relatively blue UV slope indicating recent quenching. Later, many other studies revealed the presence of similar galaxies in the epoch of reionization \citep{2024A&A...691A.310K, 2024MNRAS.529.1299V, 2025MNRAS.537..112W, 2026A&A...705A.155C, 2023MNRAS.523.3018L, 2025MNRAS.537.3662T, 2024MNRAS.52711627T}, where some of them showed the presence of a Balmer break and emission lines simultaneously. These galaxies experience temporary quenching due to feedback processes for a period of a few tens to hundreds of million years and then restart the star formation. Overall, their presence indicates the bursty nature of star formation in the early universe \citep{2024A&A...691A.310K, 2024A&A...686A.128C, 2025MNRAS.537..112W, 2026A&A...705A.155C, 2025A&A...697A..88L}.

The presence of a BB in a variety of objects, particularly at high redshifts, highlights the importance of studying the BB population as a whole in detail. However, a general study of the BB strength of galaxies, irrespective of their nature, based on JWST observations has not yet been conducted.  In this paper, we aim to understand the general trends and implications of the BB strengths of galaxies in various epochs, as observed, and to compare these with theoretical expectations. To achieve this, we use large JWST NIRCam photometry and publicly available spectroscopic datasets.

This paper is structured as follows. The observational data used in the study and the source selection methods are described in Sect.~\ref{sec_data_source_sel}. The theoretical expectations from the mock galaxy simulations are discussed in Sect.~\ref{sec_mock_sim}. The observational analysis is presented in  Sect.~\ref{sec_observation}. Section \ref{sec_discussion} discusses the general impact of the analysis on galaxy formation and evolution. Finally, our main results are summarized in  Sect.~\ref{sec_conclusion}. Unless stated otherwise, we follow the AB magnitude system and assume the $\Lambda CDM$ cosmology framework ($H_{0} = 70$ km $s^{-1} Mpc^{-1}$, $\Omega_{M} = 0.3$, $\Omega_{\Lambda} = 0.7$).

\section{Data and source selection}
\label{sec_data_source_sel}
We now describe the dataset we used for the analysis and the different steps we followed for the source selection.
\subsection{Observations}
We used the JWST observations taken as part of various public observational programs, which include the Cosmic Evolution Early Release Science (CEERS) survey \citep{2023ApJ...946L..12B, 2023ApJ...946L..13F, 2025ApJ...983L...4F}, First Reionization Epoch Spectroscopically Complete Observations \citep[FRESCO;][]{2023MNRAS.525.2864O}, The JWST Advanced Deep Extragalactic Survey \citep[JADES;][]{2026ApJS..283....6E}, and Public Release IMaging for Extragalactic Research (PRIMER, Dunlop et al., in preparation). The fields covered by these surveys are Extended Groth Strip (EGS), Great Observatories Origins Deep Survey (GOODS) - north and south, COSMOS, and Ultra-deep Survey (UDS) for CEERS, FRESCO, JADES, and PRIMER, respectively. We use the NIRCam imaging of these programs, which comprises 7 filters (F115W, F150W, F200W, F277W, F356W, F410M, and F444W), complemented by 3 Hubble Space Telescope (HST) filters (F435W, F606W, and F814W). From these images, we produced a point spread function (PSF)-matched photometric catalog. We refer to \citet{2024MNRAS.533.1808W} for details about image reduction, photometry, and catalog creation. We also use the publicly available spectroscopic observations from the DAWN JWST Archive \citep[DJA;][]{2025A&A...693A..60H} with grade 3 spectroscopic redshifts obtained until October 2025.

\subsection{Redshift estimation and initial source selection}
\label{subsec_redshift_source_sel}
The redshift and physical properties of the objects are estimated using the SED fitting tool Code Investigating GALaxy Emission \citep[CIGALE,][]{2019A&A...622A.103B}. The following assumptions are used for the redshift estimation. The star formation history is constructed using a delayed $\tau$ model with variable e-folding time and without a burst. A uniform prior ranging from 2 Myr to 13.5 Gyr on a logarithmic scale is used to constrain the e-folding time. The \citet{2003MNRAS.344.1000B} stellar population model with Salpeter Initial Mass Function \citep[IMF,][]{1955ApJ...121..161S} and Small Magellanic Cloud (SMC) metallicity, that is Z= 0.004, is used. Nebular emission is also included with zero escape fraction ($f_{esc}$) and SMC metallicity. An attenuation law with variable power-law slope, which goes from -0.5 \citep[SMC like; ][]{1981A&A....99L...5R} to 0 \citep[Calzetti like; ][]{2000ApJ...533..682C}, is considered. The redshift prior ranges from 0 to 12 with an increment of 0.05. For more details see Appendix \ref{append_cigale_para_space_sed_fits}.

\begin{figure}[htb]
    \centering
    \includegraphics[width=1\columnwidth]{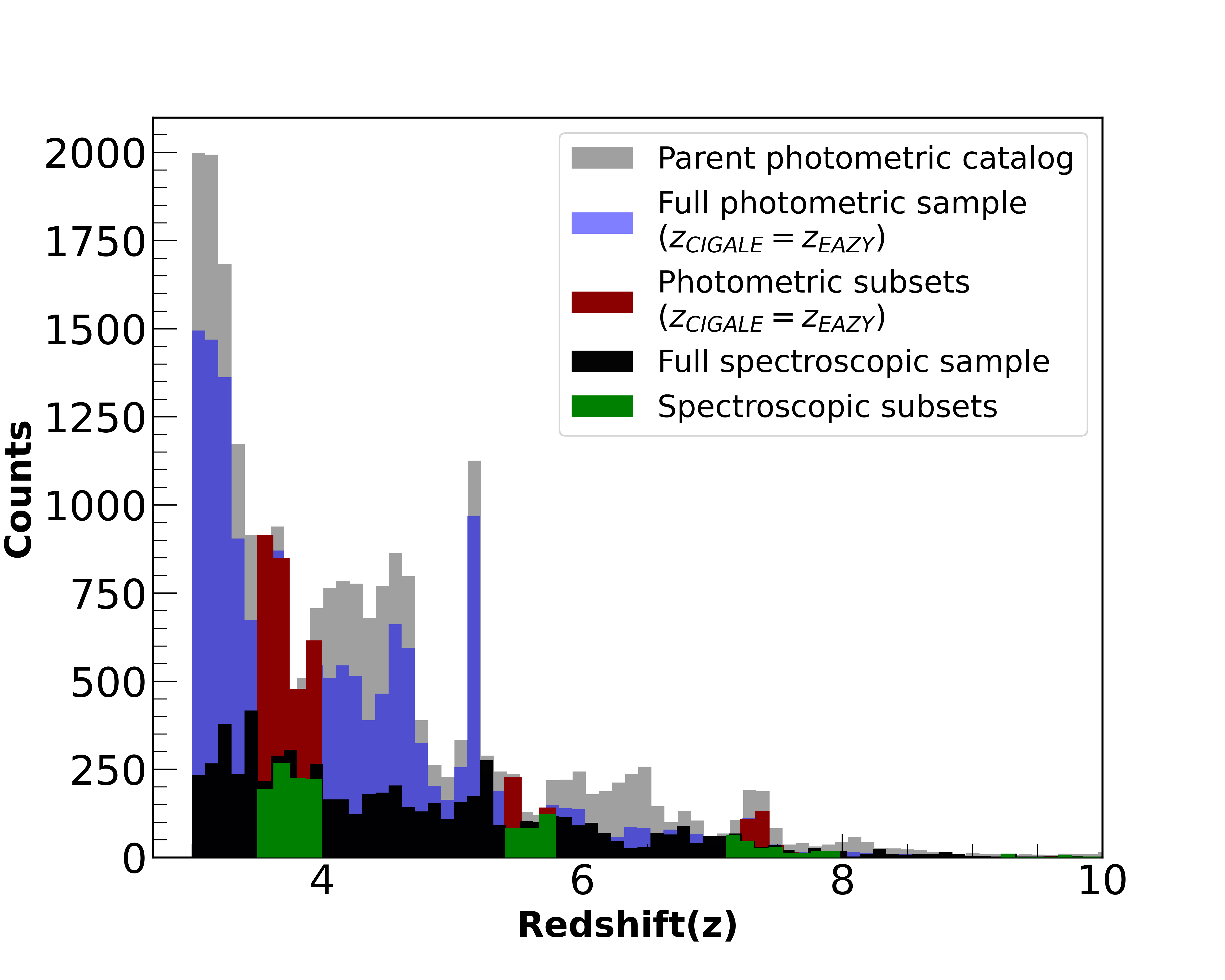}
    \caption{Redshift distribution of the various samples studied in this work. The full parent photometric catalog (which satisfies the initial selection procedures) is shown in grey, the full photometric sample with $z_{CIGALE} = z_{EAZY}$ is shown in blue, and the full spectroscopic sample is shown in black. Different subsets of both photometric and spectroscopic samples are shown in brown and green, respectively.
    }
    \label{fig_hist_z}
\end{figure}

Our photometric catalog contains 351142 objects with a stellarity flag = 0 (not a star) and 5-$\sigma$ detections in at least one of the above-mentioned JWST NIRCam filters (62444, 66519, 30208, 85095, and 106876 objects in CEERS, GOODSN, GOODSS, PRIMER--COSMOS and PRIMER--UDS fields respectively). After applying the quality flags, we select sources above $z > 2.8$ and with a narrow probability distribution of redshift estimations (sources with well-constrained photometric redshifts by requiring that the width of the posterior redshift distribution satisfies $\Delta z = z_{84} - z_{16} < 0.5$, where $z_{84}$ and $z_{16}$ denote the 16th and 84th percentiles, respectively). We refer the reader to \citet{2024MNRAS.533.1808W} for more information on the flags. Furthermore, we retain only sources for which the reduced $\chi^{2}$ value of the SED fits $\le 2.0$. These selections minimize the number of redshift interlopers in the sample. The final sample contains 16881 objects from all the fields. The redshift distribution of those objects is shown in Figure \ref{fig_hist_z}. As expected, there are more sources at lower redshift than at higher redshift. But the decrease in the number of objects from low to high redshift is not completely monotonic, and there is a peak around redshifts 4.6, 5.3, and 7.3. This can be an artefact of the SED fitting tools to find the redshift. 

We did not apply any of the above selection criteria for the spectroscopic sample since we only considered objects with grade 3 (secure redshift estimation after visual inspection). The final spectroscopic sample contains 4342 objects above redshift 3. The objects that cannot give reliable break measurements, as discussed in section \ref{subsec:bb_from_phot}, are ignored.

\subsection{Final photometric and spectroscopic samples}
\label{subsec_phot_spec_sample} 
To examine the reliability of redshift estimations, we compare the CIGALE redshift estimations from this work with EAZY redshift estimations from \citet{2024MNRAS.533.1808W}, and for a subsample of sources for which spectra are available, we compare redshift estimations from SED fitting codes with spectroscopic redshifts. This is discussed in Appendix \ref{Appedix_z_sed_fits_spec}. In general, redshift estimations from CIGALE show a good agreement with EAZY estimations and spectroscopic redshift estimations for the majority of the sample, but there are some outliers also ($\sigma_{NMAD}$ = 0.027 and $\eta_{Outlier}$ = 0.07, computed following the definitions of \citet{2008ApJ...686.1503B}). We also found that the subsample of sources for which both SED fitting codes give comparable redshifts shows a better agreement with the spectroscopic sample (see Fig. \ref{fig_z_cigale_eazy_spec}). For more details see Appendix \ref{Appedix_z_sed_fits_spec}.

To minimize redshift interlopers, we finally consider two different samples: {\em 1) } the photometric sample for which both CIGALE and EAZY give similar redshifts (that means the difference between CIGALE and EAZY redshift estimations is less than 0.2), which consists of 16881 objects, and {\em 2) } the spectroscopic sample with 4342 objects, as mentioned  above.

\subsection{BB measurement from photometry}
\label{subsec:bb_from_phot}

From our photometric and spectroscopic samples, we then selected sources in different redshift bins where we can use NIRCam filters to probe the BB. These redshift windows are designed as such: the filter redward of the break always probes 3500 -- 4861\AA\ rest frame wavelength range. This will ensure the break will always be between two consecutive filters and exclude the contamination from $\hb$ and \oiii\ emission lines longward of the break. The five selected redshift bins cover $z=$ 3.5 -- 4.0, 5.4 -- 5.8, 7.1 -- 8.0, 8.0 -- 9.2, and 9.2 -- 10. The range of these redshift windows depends on the bandwidth of the filters and the gap between two consecutive filters that are probing the break. The reliability of this method is already discussed in \citet{2024A&A...691A.310K} for redshifts $\ge 7$. They applied the above strategy to the convolved stacked spectra of \citet{2024ApJ...976..193R} and showed that the BB measurements from photometry are similar to spectroscopic ones, and the maximum difference will be 0.1 to 0.2 in flux ratio, even though the stacked spectra showed the presence of several weak lines redward of the break. They also showed that the evolution of the median of their break estimations is in agreement with the other spectroscopic studies like \citet{2024ApJ...976..193R} and \citet{Langeroodi+2024a}.

Note that only the sources that show 3 sigma detection in filters probing the BB in the particular redshift bin are considered for the  BB strength measurements and further analysis. This means that sources with upper limits in the BB-probing filters are ignored. There are 2859 (911) objects in the first bin, 441 (292) objects in the second bin, 384 (229) objects in the third bin, 79 (73) objects in the fourth bin, and 12 (27) objects in the final bin for the photometric (spectroscopic) sample. The properties of these subsamples are summarised in Table \ref{tab_BB_predict}.

\section{What BB strengths are expected?}
\label{sec_mock_sim}

\begin{table*}[tb]
    \centering
    \caption{Overview of the different redshift bins, corresponding filter combinations, and sample sizes studied (cols.~1-4). Predicted BB color-index properties (minimum and maximum values) for different star-formation histories, assuming $Z=0.004$, are listed in Cols.~5–7. The final column gives the reddening vector, $r$.
    }
    \begin{tabular}{lllllllll}
    \hline
    redshift & BB filters & $n_{\rm obj}$(Phot) & $n_{\rm obj}$(Spec) & min(BB) & max(BB)-IB & max(BB)-CSF & reddening ``vector'' $r$\\ 
                &         &         &      & [mag]   & [mag]         & [mag]      &  ($k_{\lambda1} - k_{\lambda2}$) \\
    \hline            
        $3.5-4.0$  & F150W-F200W & 2859 & 911 & -0.15 & 1.43 & 0.75 & 0.97\\
        $5.4-5.8$  & F200W-F277W & 441 & 292 & -0.09 & 1.55 & 0.65 & 1.40\\
        $7.1-8.0$  & F277W-F356W & 384 & 229 & -0.22 & 1.23 & 0.52 & 0.87\\
        $8.0-9.2$  & F277W-F410M & 79  & 73 & -0.36 & 1.66 & 0.57 &  1.87  \\
        $9.2-10.0$ & F356W-F444W & 12 & 27 & -0.16 & 1.01 & 0.42 & 0.67\\
    \end{tabular}
    \label{tab_BB_predict}
\end{table*}


We first examine the strength of the BB and its dependence on various physical properties (age, extinction, and metallicity) from the modeling perspective. To do so, we use CIGALE to generate a mock galaxy (composite stellar population) template with two different star formation histories (SFHs), that is, constant star formation (CSF) and instantaneous bursts (IB). These two SFHs probe two limiting cases of star formation histories and help us to explain the colors of most of the objects in the sample. We created mock galaxy templates with and without nebular emission for both SFHs considered here. For details on the parameter space see Table \ref{append_cigale_para_space}.

Both SFH scenarios are created using the double exponential module (shf2exp) in CIGALE. SSP models from \citet{2003MNRAS.344.1000B} are used for all computations. Except if stated otherwise, the SMC attenuation law \citep{1981A&A....99L...5R} is considered throughout the analysis, and the same attenuation law is applied for both lines and continuum\footnote{We used $"dustatt\_modified\_starburst"$ module in CIGALE for the analysis and considered powerlow slope as -0.5 and UV bump amplitude as 0 to mimic the SMC attenuation law.}. The model predictions for different SFHs and metallicities are examined using various definitions of break measurements, as discussed below.

\subsection{Different BB measurements}

\begin{figure}[tb]
    \centering
    \includegraphics[width=1\linewidth]{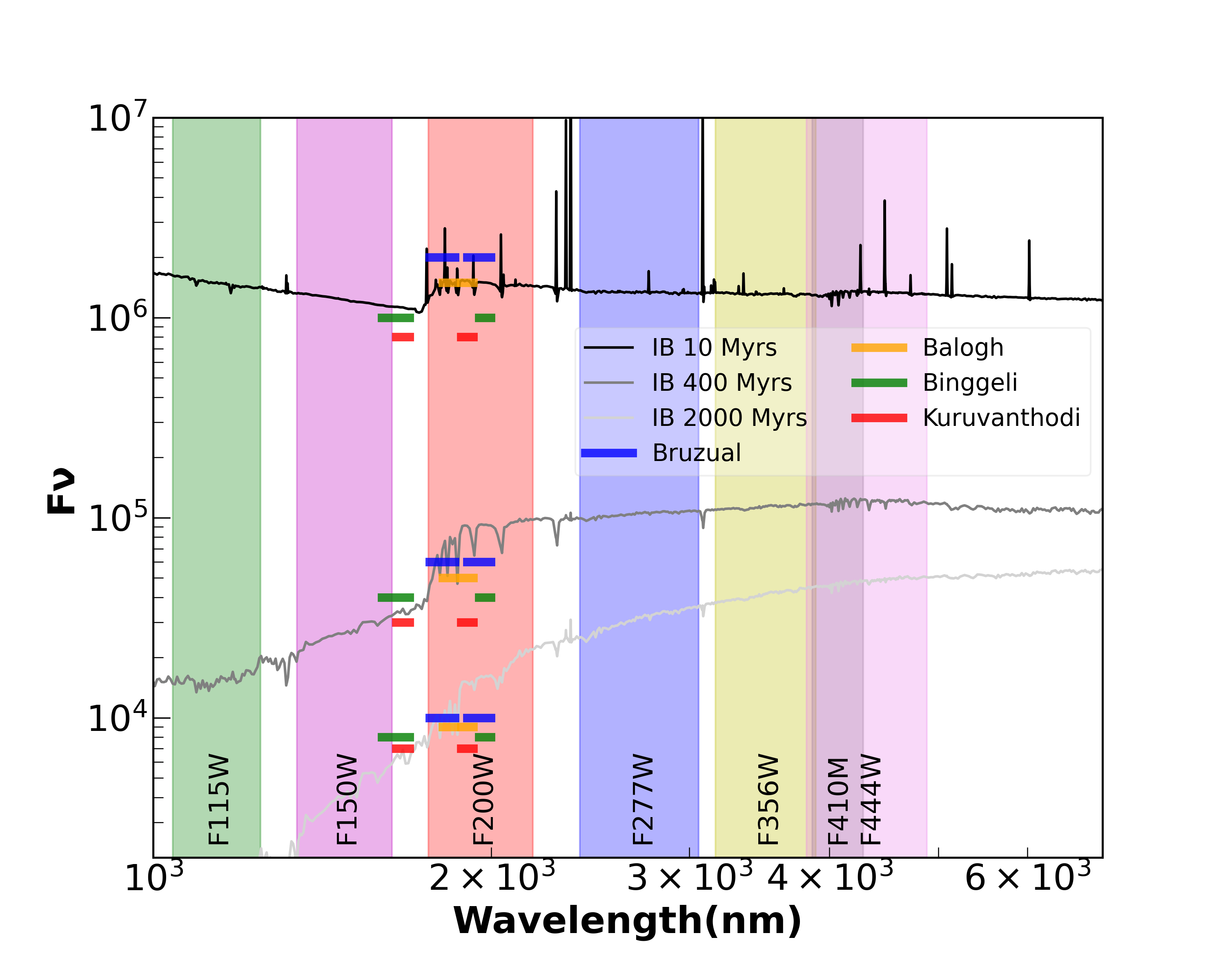}
    \caption{Model spectra for the instantaneous burst scenario for ages of 10 Myr (black), 400 Myr (grey), and 2.0 Gyr (light grey) for metallicity of 0.004 (SMC metallicity) and $z=3.7$.  The blue line indicates the D4000 break definition from \citet{1983ApJ...273..105B}, the orange line indicates the definition from \citet{1999ApJ...527...54B}, the green line indicates the definition from \citet{2019MNRAS.489.3827B}, and the red line indicates the definition used in \citet{2023A&A...674A.140K}. Various JWST filters used in this study are also shown in the figure and for this particular case ($z=3.7$), F150W and F200W filters probe the break. 
    }
    \label{fig_spec_bb_windows}
\end{figure}

\begin{figure}[htb]
    \centering
    \includegraphics[width=\columnwidth]{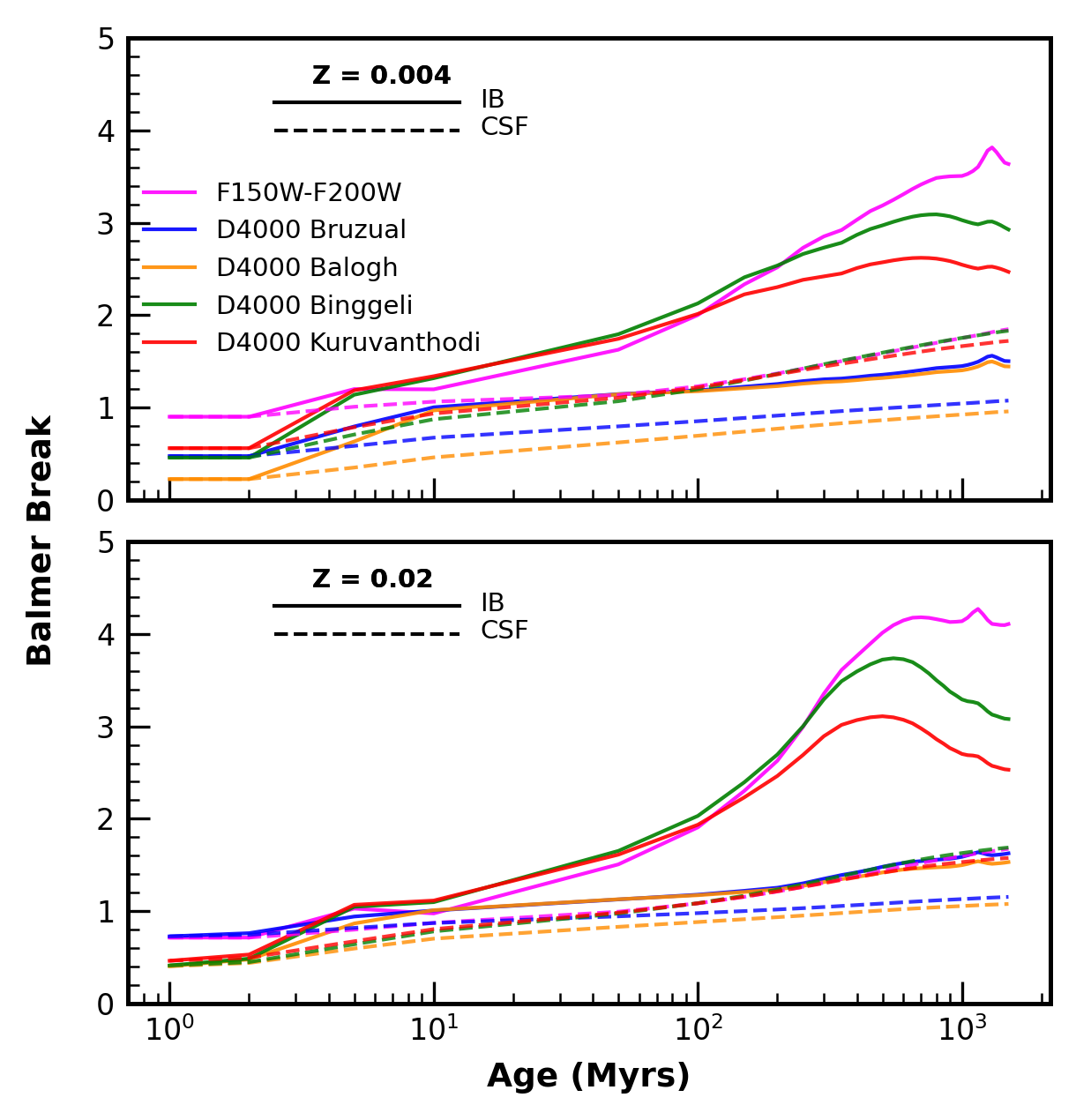}
    \caption[Predicted BB as a function of age for different metallicity ranges for the Instantaneous burst scenario]{BB estimation for metallicity Z=0.004 (\textbf{Top}) and Z=0.02 (\textbf{Bottom}) for both the IB (solid lines) and CSF (dashed lines) scenario. The blue color indicates the D4000 break definition from \citet{1983ApJ...273..105B}, the orange color indicates the definition from \citet{1999ApJ...527...54B}, the green color indicates the definition from \citet{2019MNRAS.489.3827B}, the red color indicates the definition used in \citet{2023A&A...674A.140K}, and the magenta color indicates the estimations from F150W-F200W color.}
    \label{fig_bb_diff_ind_met_inst_csf}
\end{figure}

Different measures of the BB and/or the D4000 break exist in the literature, including e.g.~from \citet[][hereafter D4000 Bruzual]{1983ApJ...273..105B}, \citet[][hereafter D4000 Balogh]{1999ApJ...527...54B}, \citet[][hereafter D4000 Binggeli]{2019MNRAS.489.3827B}, and a definition used in \citet[][hereafter D4000 Kuruvanthodi]{2023A&A...674A.140K}. All the definitions compute the BB strengths as the flux ratio (in $F_\nu$ units) redward to blueward of the break, 
$BB = \overline{F_{\nu_{3,4}}}/\overline{F_{\nu_{1,2}}}$, where $\nu_{i,i+1}$ indicates the windows used for the BB measurements. The different definitions are given in Table \ref{tab_bb_defs} and are illustrated in Fig.~\ref{fig_spec_bb_windows}. The predictions showing different BB strengths for models using different star-formation histories and metallicities are shown in Fig.~\ref{fig_bb_diff_ind_met_inst_csf} to examine and understand their behavior and the dependence on physical parameters.

\begin{table}
    \centering
    \caption{The wavelength windows used in different Balmer break measures}
    \begin{tabular}{ l | c | c }
    \hline
    \hline
    Balmer break measure & Blue continuum & Red continuum \\
                             &  (\AA)       &     (\AA) \\
    \hline
    D4000 Bruzual & 3750 -- 3950 & 4050 -- 4250 \\ 
    D4000 Balogh  & 3850 -- 3950 & 4000 -- 4100 \\
    D4000 Binggeli & 3400 -- 3600 & 4150 -- 4250 \\
    D4000 Kuruvanthodi & 3500 -- 3600 & 4000 -- 4100 \\
    \hline
    \end{tabular}
\label{tab_bb_defs}
\end{table}

Clearly, the measurements of the BB vary quite significantly depending on the wavelength domains used to probe the break. Both for instantaneous bursts and constant SFR and for all metallicities, D4000 Bingelli shows the largest dynamical range (i.e., largest relative variation with age). This is followed by the D4000 Kuruvanthodi,  D4000 Bruzual, and D4000 Balogh\footnote{D4000 Balogh is the default BB measure from CIGALE.}, which shows the weakest variations. The D4000 Bruzual and D4000 Balogh were originally proposed to probe the D4000 break in early-type galaxies rather than the BB, showing a monotonic increase with age. This is already discussed in \citet{2003MNRAS.341...33K}. 

Since such BB measurements require spectra or special filters, they will be difficult to obtain for large samples, and one may need to resort to photometry to probe the BB. Fig. \ref{fig_bb_diff_ind_met_inst_csf} shows, as an example, the BB derived from photometry using the  F150W and F200W filters, from which the F150W--F200W colors probe the BB at redshift $z \sim 3.7$. Both for IB and CSF scenarios, the BB estimations from F150W -- F200W color (from the models) are very close to the D4000 Binggeli estimations on model spectra at the given redshift. At this redshift, the F150W and F200W filter are not contaminated with strong emission and/or absorption lines and have the potential to reliably measure the break strength from photometry. Besides that, Fig. \ref{fig_bb_diff_ind_met_inst_csf} shows that the F150W-F200W color exhibits a fairly large dynamic range, which can constrain the age of the dominant population in the galaxy. Altogether, this indicates that at the right redshift intervals, the JWST NIRCam photometry is well suited to estimate the right BB strength, and we use these redshift intervals for our observational studies in the next section.


\begin{figure}[htb]
    \centering
    \includegraphics[scale = 0.35]{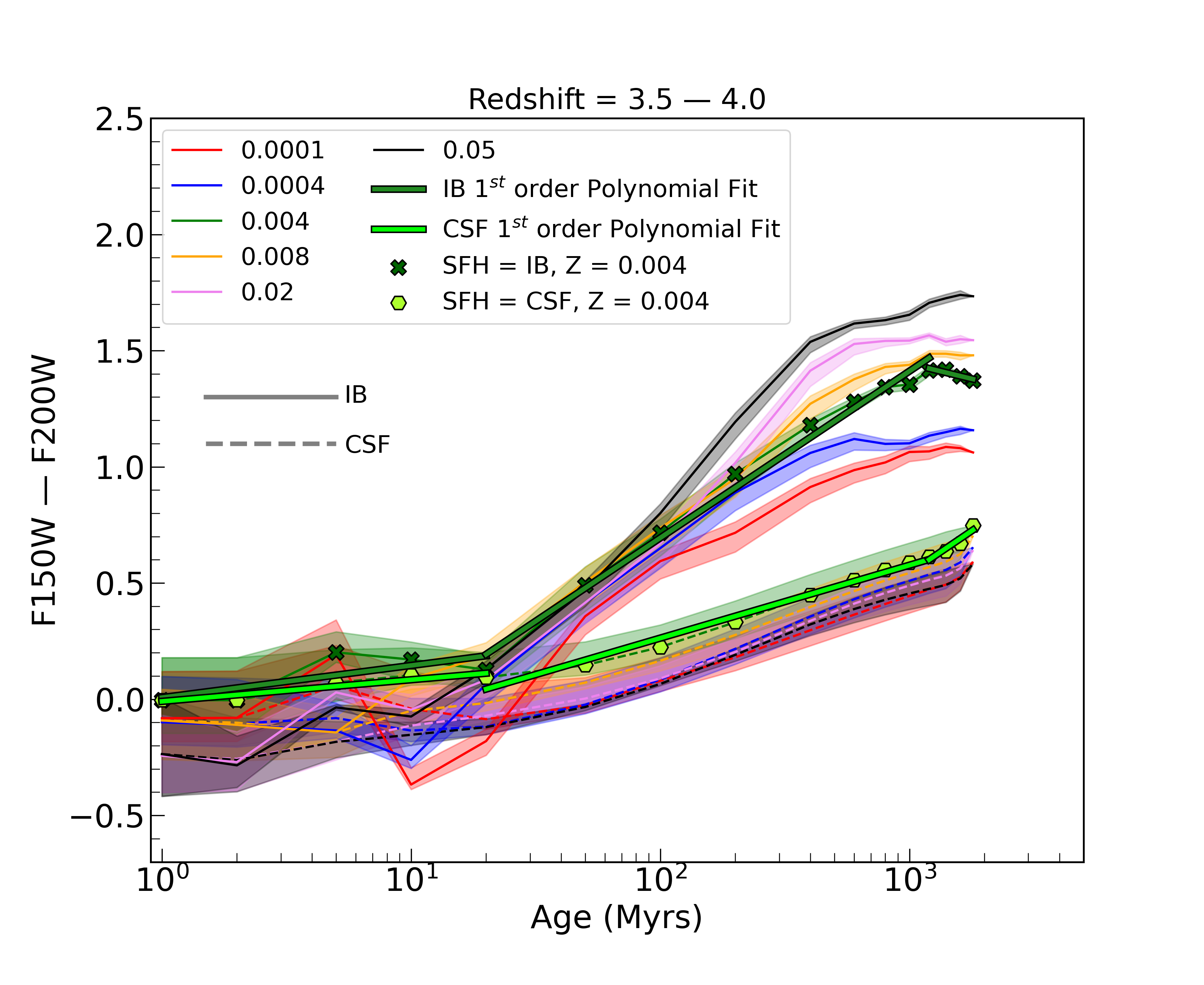}
    \caption{BB-probing color (F150W$-$F200W), derived from NIRCam filter combinations, as a function of age for different metallicities in the $z=3.5$–4.0 redshift bin, shown for the CSF and IB scenarios. The shaded region around each metallicity indicates the spread in BB due to redshift variation within the bin. The solid line represents the IB scenario, and the dashed line represents the CSF scenario. A first-order polynomial fit performed for the SMC metallicity (Z = 0.004) case is shown by the dark-green and light-green lines with black borders for the IB and CSF scenarios, respectively. The data points used for the fit are highlighted with crosses and hexagons for the IB and CSF scenarios, respectively.
    }
    \label{fig_bb_color_met_inst_csf}
\end{figure}

\subsection{Effect of age and metallicity}
\label{subsec_effect_age_met}
The age of the galaxy has a significant effect on the strength of the BB since it relates to the underlying dominant stellar population, which contributes the maximum to the integrated light. Fig. \ref{fig_bb_diff_ind_met_inst_csf} shows the evolution of different measures of the BB, with age and metallicity (Z=0.004, and Z=0.02 cases), for both the IB and CSF scenarios. See appendix \ref{fig_bb_diff_ind_met_inst_csf_p2_appendix} for Z=0.0001, 0.0004, 0.008 and 0.05 cases. Also see Fig.~\ref{fig_bb_color_met_inst_csf} and appendix \ref{fig_bb_color_met_inst_csf_z5p4_10p0} for specific filters and redshift ranges chosen for the present study.

For instantaneous bursts, the actual BB strength measured using the four most sensitive BB indicators increases with age, reaches a maximum, and then decreases again. Both the age at which the maximum is reached and the maximum BB value vary with metallicity. The reason for this behavior is simple, and due to the fact that at a given age, the integrated properties reflect the stellar population dominating at that particular age. Since A-type stars, which show the strongest BB, start dominating around a few hundred million years, the BB peaks at these ages. Subsequently, the BB decreases, and after a few  Gyrs, low mass stars start to dominate the integrated spectrum, leading again to an increase of the BB, which is then, however, due to the red slope of the SED rather than to a real spectral break.

For the highest metallicity, the BB strength will be at a maximum at $\sim 300$ Myr, while for the lowest metallicity ($\ga 1/50$ solar) this is predicted to happen around 1 Gyr. This time shift is due to the fact that the average \teff\ of metal-poor stars is higher (for the same mass and age), and, since on average stellar evolution on the main sequence proceeds towards lower \teff. It therefore takes longer for stars to reach the temperature (\teff $\sim 10^4$ K) where the Balmer break is maximum. This also explains why, at a given age, the BB is weaker at low metallicity. This overall behavior of the BB with age and metallicity is well known \citep[e.g.][]{1999A&AS..139...29G}.

For the CSF scenario (see  Fig.~\ref{fig_bb_diff_ind_met_inst_csf}), the Balmer break index (all four definitions) increases monotonically with age. Due to the constant formation of young stars, which have small (or no) BB, the BB is smaller at the same age than for instantaneous bursts. For this reason, and since it is ``averaged out'' by a mixture of different ages, the non-monotonic behavior of the BB is not seen for CSF, and the overall dynamic range is also smaller for the CSF scenario. We also see that, as in the IB case, the F150W -- F200W color correctly estimates the BB strength at the given redshift.

\subsection{Effect of nebular emission}
\label{subsec_neb_emission}
Nebular continuum emission also changes the strength of the BB at young ages for IB and throughout in the case of CSF. In this case, the BB is inverted -- it becomes a Balmer jump --  and the maximum difference in BB strength between with and without nebular emission is 0.4 (D4000 Binggeli) for ages of $\sim 1-2$ Myr. After that, this difference becomes smaller and reaches zero after $\sim 10$ Myr. For CSF, the behavior is the same at young ages, and after $\ga 10-50$ Myr, the effect of nebular emission becomes small, changing the BB by $\la 0.2$ for SMC metallicity. In measurements of the BB from photometry, the contribution of nebular emission lines  (\Oii, and higher-order lines of the Balmer series) longward of the BB can in general not be avoided \citep{schaerer&debarros2009}. Therefore, both CSF and IB models with nebular emission show a stronger break than without nebular emission. However, in the selected redshift intervals and for chosen filters the effect is small, leading to maximum BB strength differences of $\la 0.2$, both for IB and CSF.

\subsection{Effect of extinction}
\label{extinction}
Extinction strongly affects the shape of SEDs, and therefore also the photometric measures of the BB. To quantify this effect, we list in Table \ref{tab_BB_predict} the reddening vector $r$ defined by $\Delta color = r \times E(B-V) = (k_{\lambda1} - k_{\lambda2}) \times E(B-V)$ for the filters used in this work for the different redshift bins and for the SMC attenuation law. For a Calzetti law, the corresponding values of $r$ are typically 1.0 -- 1.96. This shows that for a typical extinction of $E(B-V)=0.2$ the color measuring the BB is increased by 0.17-0.28 mag, i.e., overestimated by a factor of 1.17-1.3. If the reddening (and reddening law) of the individual object is known, the true BB strength can therefore in principle easily be inferred.

\subsection{Predicted BB colors for JWST filters}

In Fig.~\ref{fig_bb_color_met_inst_csf} and Appendix \ref{fig_bb_color_met_inst_csf_z5p4_10p0} we finally show how the BB predictions translate into direct observables, i.e.~colors of the best-suited photometry available for the observations from JWST. This is done for the five redshift ranges where the BB can be probed with the selected filter sets, for the limiting cases of instantaneous bursts and constant star formation, and for all metallicities considered earlier (see Table \ref{tab_BB_predict}).

As expected from the earlier discussion, all colors show essentially a monotonic increase with age (except for the very youngest ages, i.e., < 10 Myr) and some metallicity dependence. The possible effects of dust extinction, not shown here, can easily be taken into account, as discussed above. For example, it can be seen that the color probing the BB varies thus between $\sim -0.4$ and $\sim 1.0$ mag for metallicities $\la 1/50$ solar, or up to $\sim 1.5$ mag for less extreme metallicities for simple stellar populations (instantaneous bursts). Overall, the predicted BB curves are fairly similar for all redshifts and also for all metallicity values. For $z>6$, one notices a clear decrease of the maximum BB, which is due to the age limitation by cosmology. Fig.~\ref{fig_bb_color_met_inst_csf} also shows that the effect of the redshift uncertainty within the specified window is essentially negligible.

The filters that probe the BB in different redshift bins have the potential to give a clue about the age of the underlying stellar populations (see Fig. \ref{fig_bb_color_met_inst_csf}). For the CSF scenario, the BB color index is always $< 1$ for all metallicities. For the IB scenario, this value can go above 1 after a few million years. So, if a population shows a BB color index $>1$ and zero extinction, it should not have a CSF SFH, and it might have an IB SFH with an age above a few hundred million years or a more complicated SFH (for example, CSF with an additional burst, two different burst events, etc). For the IB scenario, the maximum BB color index strength can reach $\sim 2$ for super solar metallicity at ages $>1$ Gyr in the redshift bin 5.4 to 5.8. Table \ref{tab_BB_predict} summarizes the minimum and maximum BB color indices predicted by the two SFHs for SMC metallicity in different redshift bins. The minimum BB values are identical for both SFHs within each redshift bin. However, we should note that real SFHs are complex, involving multiple bursts of varying durations and strengths. BB alone can only be used to provide an indication of the SFH or the most recent/dominant star formation event.

\subsection{Age estimation from the BB strength/BB probing colors}
To first order, it is possible to estimate the age of dominant stellar populations from the BB strength. We fitted the BB color versus age plots created using mock galaxy simulations for SMC metallicity with a first-order polynomial. With that, a slope and an intercept have been estimated for each redshift interval to measure the age of the dominant stellar populations. The age in Myr  can be calculated as: Age = $10^{m \times BB + c}$, where m and c are slope and intercept, respectively, and given in Table \ref{tab_BB_age} for IB and CSF SFH. For better age estimations, slope and intercepts are estimated for different BB strength intervals in each redshift bin. Note that, with this method, there can be larger inconsistencies for very young ages (< 5 Myr). The polynomial fit performed in the redshift interval 3.5--4.0 is shown in Fig. \ref{fig_bb_color_met_inst_csf}.

\section{Comparison with observation}
\label{sec_observation}
We now present the measurements of the BB for our large galaxy samples and investigate observationally how the BB strength varies with redshift and how it is related to observable and physical properties of the galaxies. To do so, we use objects in specific redshift bins to measure the BB from photometry, focusing on sources with reliable redshifts in five intervals between $z \sim 3.5$ and 10, listed in Table \ref{tab_BB_predict}. We also cross-matched our sample in each redshift bin with the LRD catalogs from \citet{2024ApJ...968...38K} and \citet{2025ApJ...986..126K}, finding in total 60 objects in common. These correspond to 0.1\%, 2.5\%, 3.9\%, 6.6\%, and 12.8\% of the samples in the $z=3.5$-4.0, 5.4-5.8, 7.1-8.0, 8.0-9.2, and 9.2-10.0 redshift bins, respectively.

\subsection{Distribution of BB probing colors}
\begin{figure*}[t]
    \centering
    \includegraphics[scale=0.4]{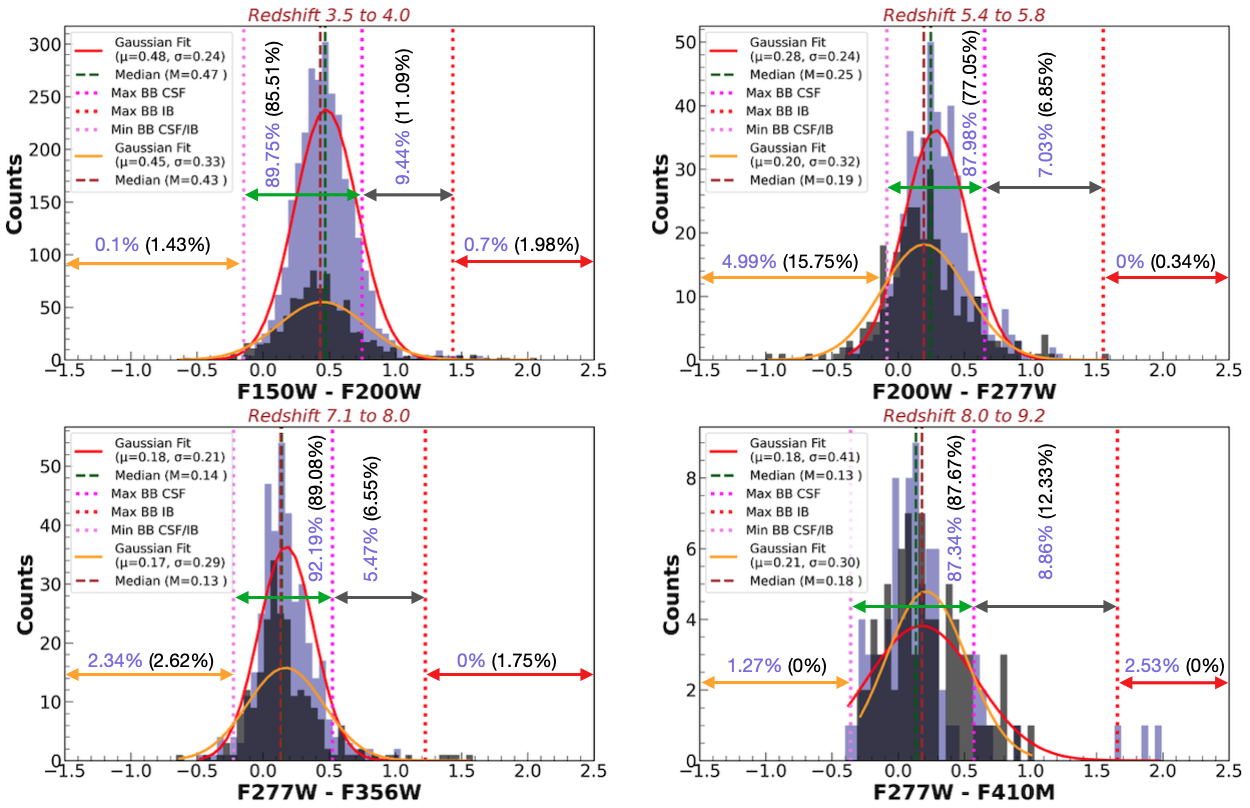}
    \caption{BB distribution of objects in different redshift bins up to $z\sim9.2$. A Gaussian fit performed for the distribution is shown in the red and yellow lines for photometric and spectroscopic samples, respectively. The median value of the distribution is shown in green and brown dashed lines for photometric and spectroscopic samples, respectively. The maximum BB expected from the CSF and IB (without extinction, E(B-V) = 0) are shown in magenta and red dotted lines, respectively. The minimum BB expected from CSF/IB is shown in a violet dashed line. The fractions of objects with BB below the minimum model prediction, consistent with the CSF scenario, consistent with only IB scenario, and inconsistent with classical models are reported for the photometric (spectroscopic) sample.
    }
    \label{fig_hist_gauss_count}
\end{figure*}

\begin{figure}[tb]
    \centering
    \includegraphics[width=\columnwidth]{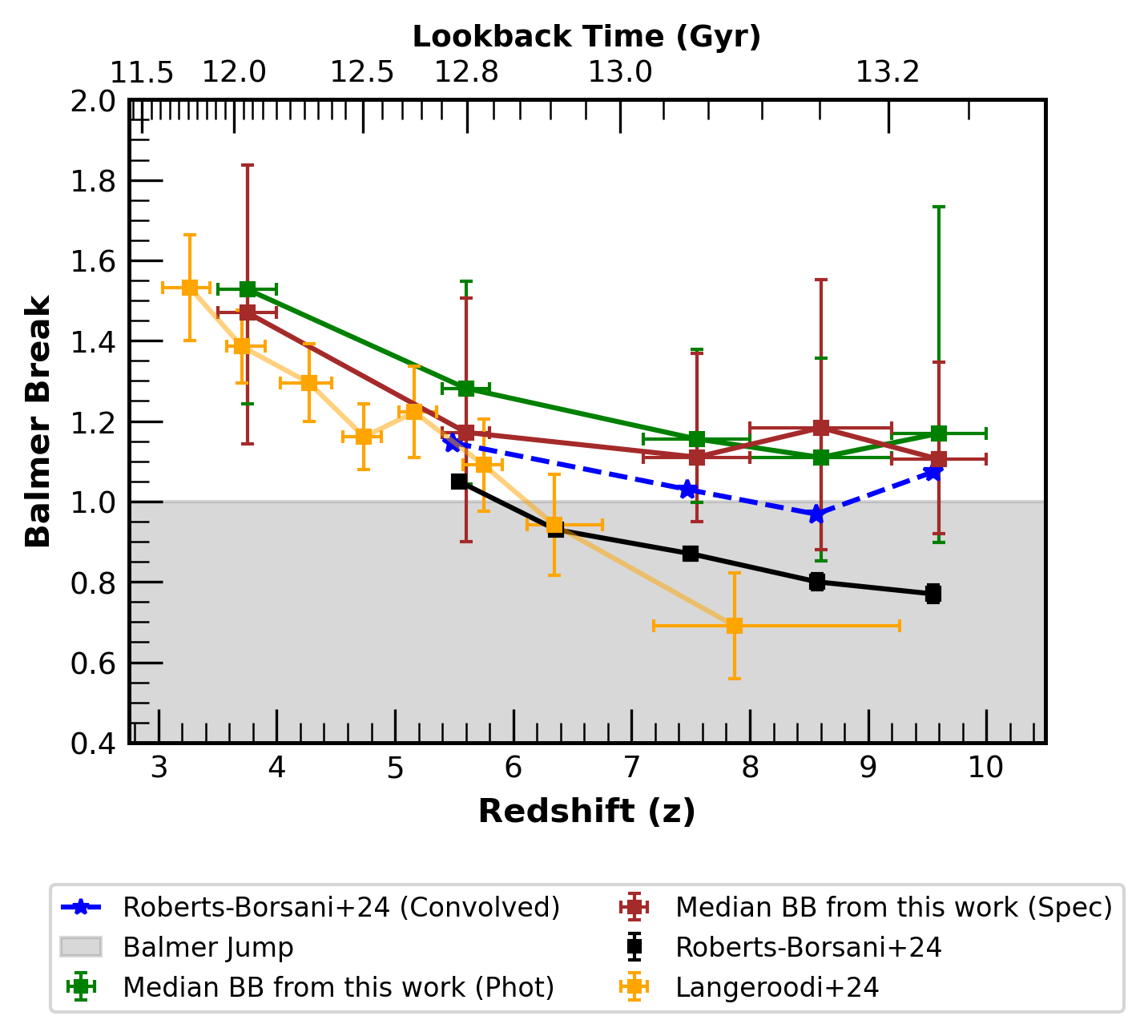}
    \caption{Median evolution of BB with redshift for photometric (Green) and spectroscopic (Brown) sample. The 16th and 84th percentiles of the distribution are shown as the error. The spectroscopic estimations from \citet{2024ApJ...976..193R} (black) and \citet{Langeroodi+2024a} (orange) are also shown. The photometric BB measurements on the stacked spectra of \citet{2024ApJ...976..193R} are shown in a blue dashed line. The shaded grey region highlights the Balmer jump.
    }
    \label{fig_bb_vs_z}
\end{figure}

The observed distribution of the BB color in each redshift bin is shown in Fig.~\ref{fig_hist_gauss_count} and Appendix \ref{fig_hist_gauss_count_part2}, and the evolution of the median of the same quantity in Fig.~\ref{fig_bb_vs_z}. At all redshifts, the BB distribution is approximately Gaussian with a width of $\sigma \approx 0.25$ mag. At the lowest redshifts ($z=3.5-4$), the median and mean of the BB distribution is $\sim 0.5$ mag, after which it decreases towards higher redshift. This is corroborated by a Kolmogorov–Smirnov (KS) test, which indicates that the BB distributions of each redshift bin are statistically distinct. The KS statistic (D) exceeds 0.2 for all pairwise comparisons, with p-values below 0.05 in all cases except for the highest-redshift bin, where the lack of significance is possibly due to limited statistical power. To first order, i.e.~neglecting metallicity and dust attenuation variations, we interpret the observed shift of the BB distribution  as being due to a decrease in the mean age of stellar populations with increasing redshift, which we quantify below.

Minimum and maximum BB strengths predicted from both IB and CSF models are tabulated in Table \ref{tab_BB_predict} and overplotted in Fig.~\ref{fig_hist_gauss_count}. Except for a few galaxies, which we discuss below, all observed BB values are within the range of model predictions. This also includes negative values of the BB color, i.e.~Balmer jump, which are presumably due to free-bound nebular continuum emission of hydrogen, as seen in metal-poor strongly star-forming low$-z$ galaxies \citep[e.g.][]{2024MNRAS.528L..10I}, and measured in stacks of JWST galaxy spectra at $z>6$ \citep{2024ApJ...976..193R}.

The high-end tail of BB requires star-formation histories that are not constant, and requires declining or instantaneous bursts in the most extreme scenario. This is typically the case for $\sim 10$\% of the objects, as indicated in Fig.~\ref{fig_hist_gauss_count}. The fraction of objects that need IB scenario to explain the BB strength decreases from $z=3.5$ to $z=8.0$. This likely reflects the gradual absence of quiescent galaxies (QGs) with increasing redshift. 
However, the fraction of objects requiring an IB scenario increases again from $z=8.0$ to $z=10$, possibly due to LRDs this time. We also speculate that the high proportion of objects with BB greater than the maximum CSF value in the highest redshift bin is due to low statistics and possible interlopers. 

Considering objects with a ``very strong'' BB (say BB$\ge 2.5$ in $F_\nu$ units, or a BB color $>1$ in magnitudes), we find that they are rare but found across all redshift bins in the sample, as also shown in Fig.~\ref{fig_bb_vs_z_lit}. However, these strong BB sources at $z<5.8$ and $z>7$ are two distinct populations, as inspection of their SEDs show. At $z\leq 5.8$ a significant number of strong BB sources are quiescent or post-starburst galaxies with possible extinction, whereas the strong BB sources at $z>7.0$ are LRDs. Furthermore, the BB of some LRDs is significantly stronger than the maximum predicted by theoretical models (IB scenario).
\citet{2025arXiv251110725H} indicated that the LRD fraction shows a decreasing trend from redshift 3 to 8 and again increases from redshift 8-9. We might be observing a similar trend in terms of BB strength, as they are some of the strong BB-producing objects. Possibly, the redshift $7.1-8.0$ bin indicates the transition of the dominant BB-producing sources. This indicates that even though the age, dust, and/or metallicity primarily govern the redshift evolution of the BB strength (like the median values of BB) of the majority of galaxies, the presence of LRDs significantly influences the BB distribution at $z>8$ cases, as they are the only possible strong BB sources at those redshifts. This effect can be observed as a higher standard deviation in the $z=8-9.2$ bin. Objects with strong BB are also discussed in Sect.~\ref{sec_discussion}.

At $z \sim 3.5-4$ there are a few objects for which the BB strength exceeds the maximum value predicted by the IB scenario. After inspecting the SED fits of the sources, we conclude that the majority of these sources are dusty galaxies with an extinction $E(B-V) \approx 0.2$ or above. Some are also LRDs, e.g.~the well-known extreme BB object at $z \sim 3.5$ \citep[the "Cliff",][]{2025A&A...701A.168D}. Similarly, there are also objects whose BB strength is less than (i.e., bluer than) the lowest value predicted by both SFH scenarios. This is due to the fact that photometry has limitations to measure the true Balmer jump values due to emission lines, as already discussed in section \ref{subsec_neb_emission}.

In Fig.\ref{fig_bb_vs_z}, we compare the median BB strength evolution of our photometric and spectroscopic sample with the spectroscopic studies from \citet{2024ApJ...976..193R} and \citet{Langeroodi+2024a}. The median BB strength at redshift $z\sim3.5$ is around 1.54 (1.49) for the photometric (spectroscopic) sample in flux ratio and decreases as redshift increases to 1.11(1.09). These estimations are in agreement with \citet{Langeroodi+2024a} to $z\sim6$, as the average population shows a BB rather than a Balmer Jump. However, for $z>6$, both spectroscopic studies show a Balmer jump instead of a break for the average population. However, our photometric method slightly overestimates the BB strength at $z>6$ as weak emission lines contribute in some cases (see Sect.~\ref{subsec_neb_emission}). To understand this, we convolved the stacked spectra of \citet{2024ApJ...976..193R} to the filters we use in the respective redshift intervals, and our median values at all redshifts are in agreement with these convolved values (See Fig. \ref{fig_bb_vs_z}). We note also that our median values may be slightly higher than literature values at z=8-9.2 due to LRDs, since  the LRD fraction is higher in this bin and that some LRDs have a strong BB. Excluding the known LRDs reduces the median values by 0.04-0.05 in this redshift bin. In any case, the studies of \citet{2024ApJ...976..193R} and \citet{Langeroodi+2024a} are subject to selection effects, which are different than those of our photometric study. Overall, we cannot rule out the possibility of a decreasing trend in BB from $z\sim4$ to $z\sim8$, followed by a period of relative constancy or slight increase till $z\sim10$ due to LRDs. Large, unbiased spectroscopic surveys that minimize selection effects are needed to establish this.

\subsection{Age distribution of the sources}

To estimate the stellar population ages of the galaxies, we use the above model predictions of the BB for both IB and CSF, assuming a typical metallicity $Z=0.004$, and neglecting extinction. The resulting age distributions are shown in Fig.~\ref{fig_hist_age_inst_csf} and \ref{fig_hist_age_inst_csf_spec} in the appendix, for all redshift bins. As expected, ages depend strongly on the assumed star-formation histories, and our absolute values are therefore only indicative. In both cases and for all redshifts, the distributions peak at ages below $\sim100$ Myr, except for the lowest redshift bin. For the IB scenario, the median age ranges from 48 (42) Myr to 8 (6) Myr from the lowest to highest redshift bin for the photometric (spectroscopic) sample. The fraction of objects with ages below 100 Myr ranges from $\approx 85\%\ (85\%)$  to $\approx 98\%\ (97\%)$ for the photometric (spectroscopic) sample.

As expected, the CSF scenario provides higher age estimations, with the median age decreasing from 350 (235) Myr to 20 (9) Myr from $z \sim 3.5$ to $z \sim 10$ for the photometric (spectroscopic) sample. In this scenario, the fraction of objects with ages $<100$ Myr increases from $\approx 16\%$ ($26\%$) to $\approx 70\%$ ($74\%$) from the lowest 3.5 to 8.0 - 9.2 (3.5 to 9.2 - 10.0) bins for the photometric (spectroscopic) sample. As shown already above, the CSF scenario cannot explain a fraction (typically $\sim 7-19$\%) of the objects with ``strong'' BB, whose value exceeds that predicted for populations reaching the age of the Universe at the considered redshift. Such unrealistic age estimations for some of the very strong BB sources in the highest redshift bins are not included in the histogram given in Figs.~\ref{fig_hist_age_inst_csf} and \ref{fig_hist_age_inst_csf_spec}. The existence of these galaxies implies rapidly decreasing (or quenched) SFHs and/or unusually high dust attenuation.

Overall, in both scenarios, the fraction of objects with ages $ <100$ Myr increases and the median age decreases as redshift increases, which indicates an increasing fraction of young galaxies at high redshift. However, the exact fraction of young and old objects depends on the exact SFHs of these galaxies, which is not constrained by our analysis.

\subsection{Does the Balmer break strength correlate with physical parameters?}

\begin{figure*}[htb!]
    \centering
    \includegraphics[width=1\linewidth]{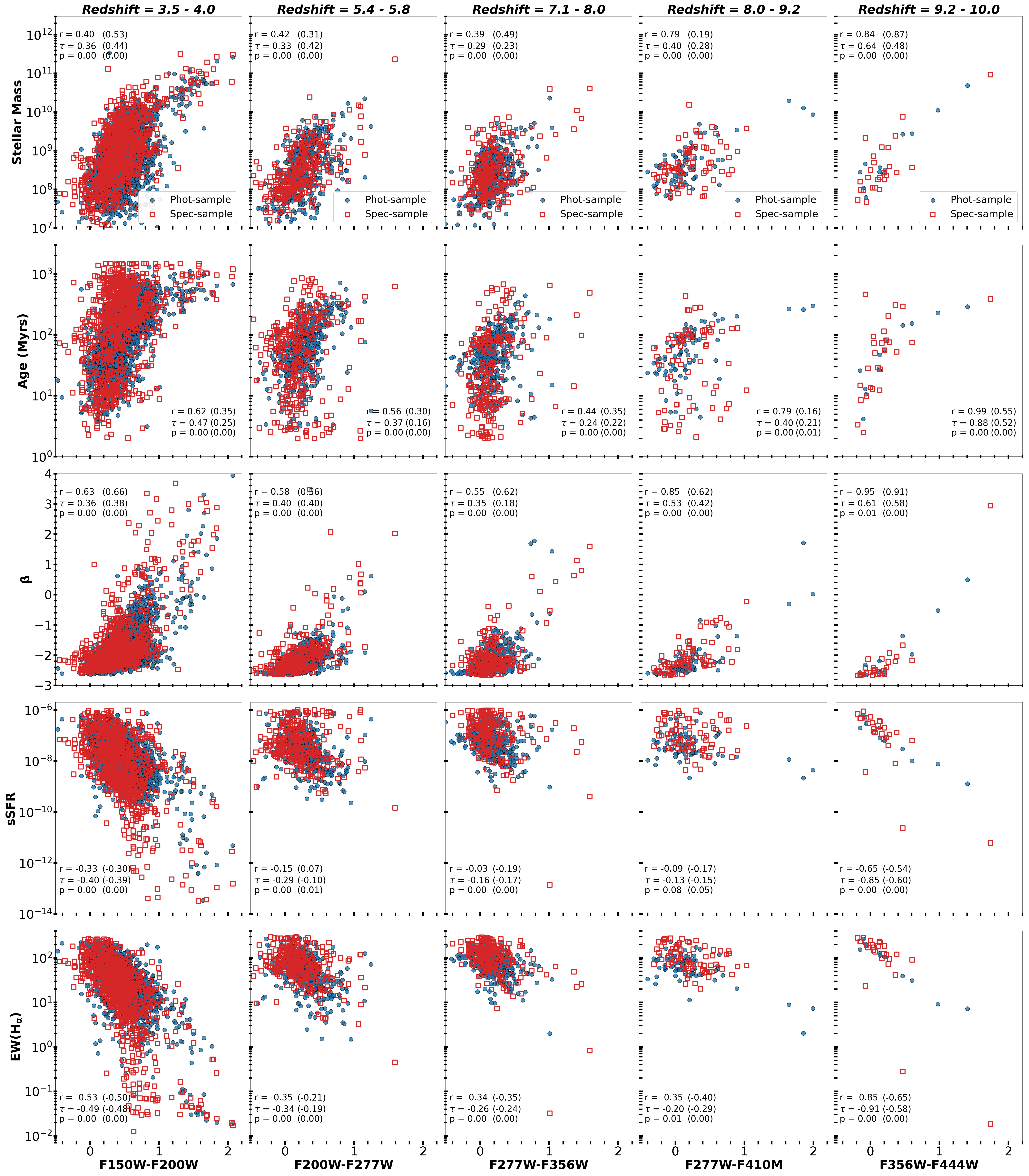}
    \caption{Correlation of Balmer break strength with various physical parameters for the photometric and spectroscopic sample. The blue dot represents the photometric sample, while the red squares represent the spectroscopic sample. Redshift increases from left to right (in the order of 3.5-4, 5.4-5.8, 7.1-8.0, 8.0-9.2, and 9.2-10). Each row corresponds to various physical parameters. From top to bottom, it indicates Stellar mass, Age, and $\beta$ slope, sSFR, and EW$(H_{\alpha})$. Within each plot, Pearson and Kendall correlation coefficients are given for photometric (spectroscopic) sample.
    }
    \label{fig_param_vs_bb_pear_kend}
\end{figure*}

In this section, we investigate how the observed BB color index correlates with various physical parameters estimated from the SED fits. The following physical parameters are considered: stellar mass, age of the object (dominant stellar population), UV slope ($\beta$), specific star formation rate (sSFR), $EW(H_{{\alpha}})$, $EW(H_{{\beta}})$, extinction ($E(B-V)$), star formation rate (SFR), absolute V band magnitude ($M_{V}$), and absolute UV magnitude ($M_{UV}$). The results are shown in Fig.~\ref{fig_param_vs_bb_pear_kend} and appendix \ref{fig_param_vs_bb_pear_kend_p2} for both the photometric and spectroscopic sample. The Pearson and Kendall correlation coefficients are also mentioned in the figure insets.

In general, we find a moderate to strong correlation of the BB color with mass, age, E(B-V), and $\beta$ slope for all redshift bins. For mass, the redshift bins between 3.5 and 8.0 show a moderate correlation; however, the highest redshift ones show a strong correlation. A  moderate correlation between D4000 break strength and stellar mass has been reported with low redshift galaxies in the Sloan Digital Sky Survey \citep[SDSS;][]{2003MNRAS.341...54K} and intermediate redshift ($0.6<z<1.2$) galaxies in the HST PEARS Survey \citep{2018ApJ...867..118K}. In the latter case, for galaxies with stellar masses greater than $10^{9.44}\msun$, they report a Pearson correlation coefficient of 0.44 for the entire sample and 0.48 for the subsample of spheroidal galaxies. Although these values are comparable to our estimations in the redshift bins at z = 3.5–8, we caution that our sample primarily probes the Balmer break rather than the D4000 break, and that we probe an even lower mass end. A relatively weak or no correlation between BB strength and stellar mass has been reported for $z=5-10$ simulated galaxies in the First Light and Reionisation Epoch Simulations \citep[FLARES;][]{2024MNRAS.527.7965W} and FIRE-2 simulations at $z=7-9$ \citep{2019MNRAS.489.3827B}. 
Our observations also show a broader distribution of the BB strengths than found in the FIRE-2 simulations.

As already shown in Sect.~\ref{subsec_effect_age_met}, age is a primary driver for the BB strength and the observations also show moderate to strong correlations between BB strength and age. The lowest redshift bin shows a strong correlation, then the correlation coefficient decreases with redshift up to the 7.1-8.0 bin, and then again increases to very strong correlation in the last two redshift bins. The highest redshift bins show the strongest correlation with age, possibly as the age can be the primary contributor to BB strength at high redshift. The $\beta$ slope and the BB are two independent indicators of stellar population age and star-formation history. Both generally increase as the star-formation activity declines, although $\beta$ slope is more strongly affected by dust attenuation. This is also corroborated by a similar trend observed for the $\beta$ slope also, as the highest redshift bins show the strongest correlation between BB strength and $\beta$ slope.  A moderate to strong correlation is observed between BB strength and $\beta$ slope in $z=3.5-8.0$ redshift bins. In comparison, the FLARES simulations shows a weak-moderate correlation for BB strength with age (Pearson coefficient = 0.37) for the simulated galaxies at $z=5-10$ \citep{2024MNRAS.527.7965W}. 

Our data also show a moderate to very strong correlation between BB strength and extinction. Furthermore, the correlation between BB and E(B-V) appears to become stronger with increasing redshift, which could be due to LRDs, as our SED fits require significant extinction to reproduce very high BB strengths. A moderate correlation (Pearson coefficient = 0.52) between BB strength and extinction is observed in simulated galaxies in FLARES \citep{2024MNRAS.527.7965W}.

A weak or moderate to strong negative correlation of the BB with sSFR, $EW(\ha)$, and $EW(\hb)$ is evident in some redshift bins. \citet{2024MNRAS.527.7965W} observed a very strong negative correlation between BB strength and sSFR for the simulated galaxies. However, we find a moderate negative correlation of BB strength with sSFR in the lowest redshift bin (3.5-4.0) and strong negative correlation in the highest redshift bin (9.2-10). Other redshift bins show no or very weak negative correlation between BB strength and sSFR. $EW(\ha)$ shows a moderate (Pearson’s r = -0.53) negative correlation with BB strength in the lowest redshift bin, strong (Pearson’s r = -0.85) negative correlation in the highest redshift bin, and a weak (Pearson’s r = -0.34/-0.35) negative correlation in other redshift bins. A similar trend for $EW(\ha)$ with BB strength has been reported by \citet{2026A&A...705A.155C} for their spectroscopic sample at $z=5-7.4$. We also report a similar negative correlation between BB strength and $EW(\hb)$, with a moderate correlation (Pearson's r = -0.39 to -0.55) in z < 9.2 redshift bins and a very strong negative correlation in the highest redshift bin (Pearson's r = -0.84). Our results indicate that both $EW(\ha)$ and $EW(\hb)$ are valuable diagnostics, comparable to sSFR, for characterizing the galaxy population at $z=3.5$–10. Similar to the BB, both EW(H$\alpha$) and EW(H$\beta$) are indicators of star formation activity and the dominant stellar populations within galaxies. As star formation declines, EW(H$\alpha$) and EW(H$\beta$) decrease due to the rapid disappearance of massive O- and B-type stars and the corresponding weakening of nebular emission. At later times, EW(H$\alpha$) and EW(H$\beta$) may appear in absorption owing to the increasing contribution of A-type stars. The observed anti-correlation between BB strength and EW(H$\alpha$) thus arises from the evolution of the underlying stellar population.

The observations show no correlation of the BB with SFR, at all redshifts considered. Apart from the highest redshift bins, the BB also does not significantly correlate with the absolute UV or V band magnitudes. This is consistent with the results presented in \citet{2024A&A...691A.310K}, as the highest redshift bins might be affected by poor statistics, thereby producing strong correlations. However, a strong anti-correlation with BB strength and UV luminosity has been reported for FLARES galaxies \citet{2024MNRAS.527.7965W}, and a similar correlation has been observed with BB strength and $M_{UV}$ for FIRE-2 galaxies \citep{2019MNRAS.489.3827B}.

\section{Discussion}
\label{sec_discussion}
\subsection{Comparison with other studies in the literature}
\begin{figure*}[!ht]
    \centering
    \includegraphics[width=18cm]{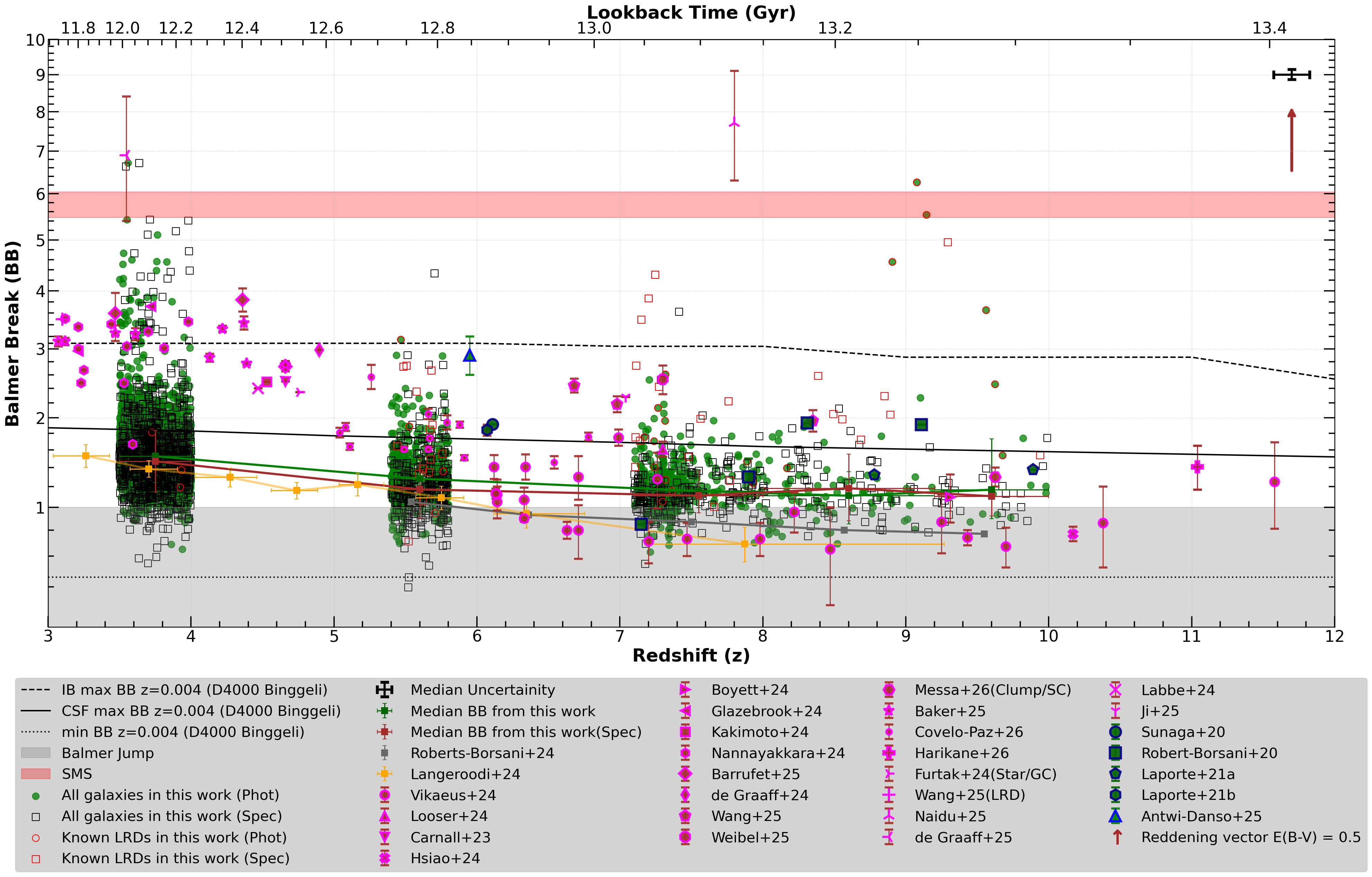}
    \caption{Comparison of the Balmer break sources in this work with the literature. Small green solid circles (photometric sample) and black squares (spectroscopic sample) indicate the sources in this work. The Black dashed line indicates the maximum BB strength expected from IB for metallicity Z=0.004, the black line indicates the maximum BB strength expected from CSF at Z=0.004, and the dotted black line indicates the minimum BB expected from both IB and CSF (Both IB and CSF produce the same break strength at younger ages). Note that the minimum and maximum BB strengths shown are derived from the D4000 Binggeli definition. The BB measurements of the stacked spectra from \citet{2024ApJ...976..193R} and \citet{Langeroodi+2024a} are shown in grey and orange squares, respectively. The median BB strengths of the photometric and spectroscopic samples from this study are indicated by the green and brown lines, respectively. Both the redshift and Balmer break strength are estimated from the spectra for the sources from \citet{2024MNRAS.529.1299V, 2024Natur.629...53L, 2023Natur.619..716C, 2024NatAs...8..657B, 2024MNRAS.527L...7F, Glazebrook+2024a, 2024ApJ...963...49K, 2024NatSR..14.3724N, 2024ApJ...969L..13W, 2024ApJ...973....8H, 2024arXiv241204557L, 2025ApJ...984..121W, 2025MNRAS.537.3453B, 2025NatAs...9..280D, 2025ApJ...983...11W, 2025A&A...702A.270B, 2025arXiv250316596N, 2025A&A...701A.168D, 2025MNRAS.544.3900J, 2026A&A...705A.173M, 2026A&A...705A.155C} and \citet{2026arXiv260121833H}. Objects in this group are shown in various symbols with magenta outlines and brown filling. While the redshift is estimated from spectroscopy but the BB strength is measured from the photometry for the objects from  \citet{2020MNRAS.497.3440R, 2020IAUS..341..309S, 2021MNRAS.505.3336L} and \citet{2021MNRAS.505.4838L}, and these objects are shown in dark blue outline with green filling. A photometric candidate reported by \citet{2025arXiv251203154A} from the CANUCS medium band survey is shown in triangle with light blue outline and green filling. Note that the Y axis is shown in Square-Root Scaling.
    }
    \label{fig_bb_vs_z_lit}
\end{figure*}

We now compare the BB measurements from our large galaxy sample to others reported in the literature. Figure \ref{fig_bb_vs_z_lit} shows such a comparison, including literature data with BB measurements both from spectroscopy and photometry. At first glance, our measurements and the literature data agree overall and show a similar redshift evolution of the BB. A substantial number of strong BB objects ($BB>3.1$) emerge at $z\lesssim3$–4. The strongest BB among both the photometrically and spectroscopically selected sources is the same object ( RA:214.8538991 Dec:52.861365), with a break strength of $\sim6.7$ in flux ratio at $z\sim3.6$. Our SED fits suggests that this object is dust-obscured with a BB. However, the available spectra do not cover the BB. Future spectroscopic observations covering the break is required to establish the real nature of this source\footnote{This object is identified as a highly extincted quiescent galaxy (ID: 3D-EGS-27584) in \citet{2018A&A...618A..85S}}. As noted above, the strongest BB candidate is reported by \citet{2025arXiv250316596N} at $z=7.8$, with a BB strength of 7.7 (flux ratio). Our sample suggests that similarly extreme BB galaxies are present across a wide redshift range. The spectroscopically selected quiescent galaxy with the strongest BB was reported by \citet{2025MNRAS.537.3453B}, with $BB\sim3.8$ at $z=4.36$. All these strong BB objects are above the maximum BB expected from IB, requiring additional extinction and/or higher metallicity to explain the break strength in the classical way. Notably, these objects begin to appear at $z \la 4$, near the epoch of cosmic noon, where dusty star-forming galaxies (DSFGs) are also prevalent. Some of the sources at z>6 with $BB > 3$ are known LRDs.

Figure \ref{fig_bb_vs_z_lit} also shows that the majority of objects in both the sample and in the literature can be explained by standard stellar population models. At $z<5$, most literature sources have a BB exceeding the strength predicted for CSF models, which therefore require a SFH which rapidly declined (similar to an IB) to reproduce their break strength. Alternatively they could have significant extinction. The reddening vector corresponding to E(B-V) = 0.5, as shown in Fig. \ref{fig_bb_vs_z_lit}, indicates the extent to which extinction can contribute to BB strength. Clearly, an extinction of $E(B-V)=0.5$ with the IB SFH is sufficient to explain the break strengths of the majority of sources at $z=3.5-4.0$. Our SED modeling of these sources indicates significant extinction (E(B-V) > 0.1) and evolved stellar populations with ages in the range 400–1400 Myr. This suggests a significant dust content in early post-starburst galaxies (PSBs) and quiescent galaxies (QGs), consistent with the findings of \citet{2024ApJ...974..145S} and \citet{2025ApJ...985..125S}.

An alternative, though non-standard, explanation for such strong Balmer break strengths could be the presence and contribution of cool supermassive stars \citep[SMS;][] {2020A&A...633A...9M, 2023A&A...674A.140K}. SMSs could, e.g., form during the early stages of the formation of a proto-globular cluster \citep{2018MNRAS.478.2461G} and produce the strongest stellar-origin BB \citep[see Fig.~10 in ][]{2023A&A...674A.140K}. We show the BB strength of cool SMS models from \citet{2020A&A...633A...9M} in Fig. \ref{fig_bb_vs_z_lit}. These SMS models predict BB $\sim 5.5-6$ (for metallicities $Z=0.0004-0.02$), which, with or without extinction, can potentially explain the BB strength of LRDs, even the strongest BB from \citet{2025arXiv250316596N}. The potential of SMS to reproduce LRD spectra has previously been discussed by \citet{2025A&A...701A.168D, 2025arXiv250712618N,Chisholm2026Little-Red-Dots}.

\subsection{Interpretation of BBs -- constraints on SFHs and formation redshifts} 
\begin{figure*}[!ht]
    \centering
    \includegraphics[width=18cm]{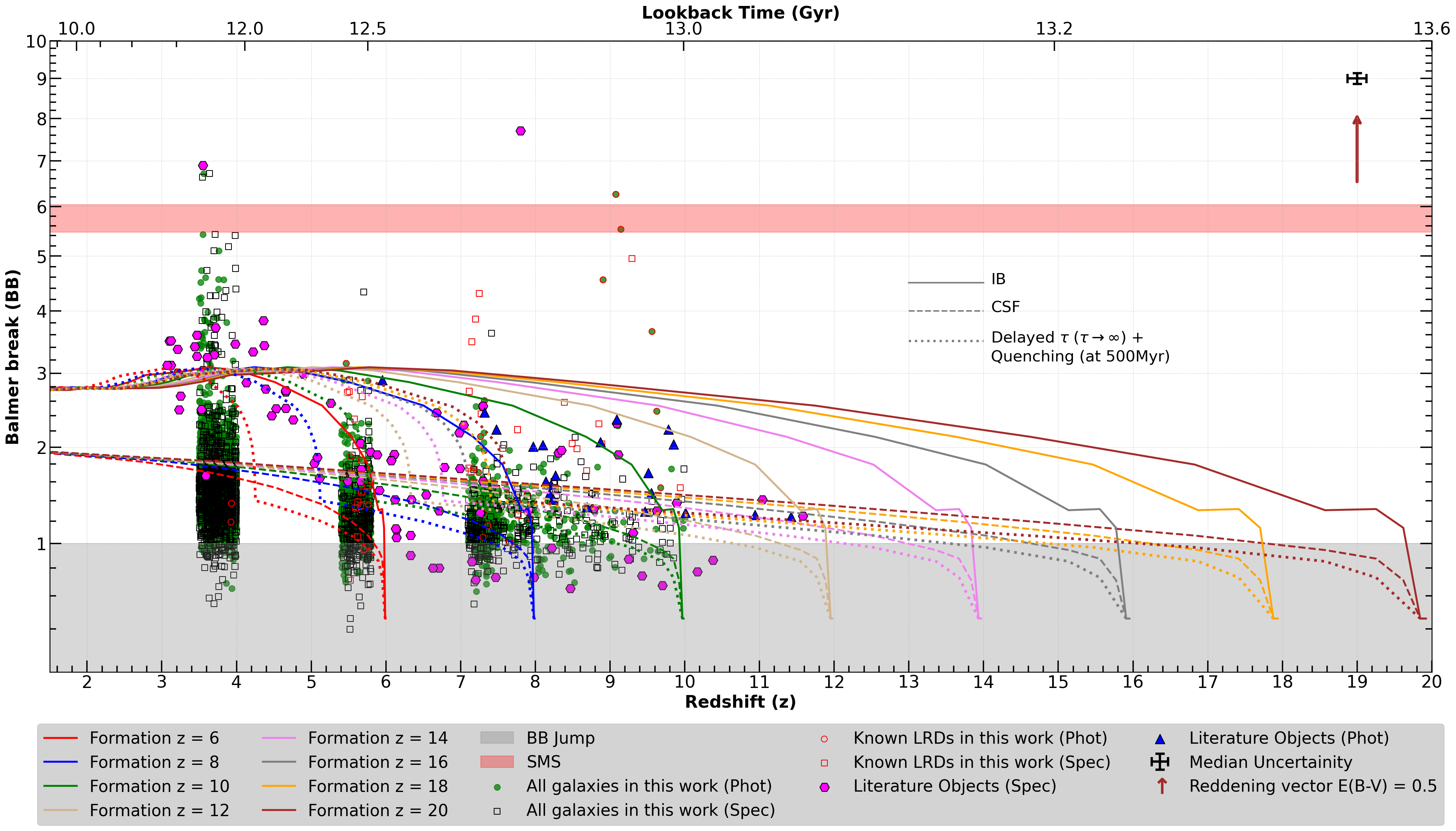}
    \caption{Evolution of Balmer break strength with formation redshift. The solid line indicates IB SFH, the dashed line indicates CSF SFH, and the dotted line indicates a delayed-$\tau$ model with a very large $\tau$ value and quenched after 500 Myr. Different colors indicate different formation redshifts. The sources shown in Fig. \ref{fig_bb_vs_z_lit} are also overplotted. The photometric and spectroscopic samples in this study are indicated as green-filled circles and black open squares, respectively. Spectroscopic sources from the literature are shown as magenta stars, while selected photometric sources at $z\gtrsim6$ are shown as blue triangles. The photometric sample is drawn from \citet{2023Natur.616..266L, 2023MNRAS.523.3018L, 2023MNRAS.519.3064F, 2023MNRAS.518.4755A, 2025arXiv251203154A}. Note that the Y axis is shown in Square-Root Scaling. 
    }
    \label{fig_bb_vs_form_z_lit}
\end{figure*}

We now investigate the potential of the BB to constrain the SFH and formation redshift of galaxies in a simple fashion. To do so, we map the age evolution of the BB strength as a function of redshift for the two SFHs considered here (IB and CSF), and for different assumptions on the formation redshift (i.e.~the start of star formation). The result is shown in Fig. \ref{fig_bb_vs_form_z_lit}, where we also include a delayed-$\tau$ model with a very large $\tau$ value and quenching occurring after 500 Myr. We present the age evolution of the BB for formation redshifts spanning from $z=20$ (180 Myr after the Big Bang) to $z=6$ (950 Myr after the Big Bang), in steps of $\Delta z=2$. We also overplot the observed data points already shown in Fig. \ref{fig_bb_vs_z_lit}.

Fig. \ref{fig_bb_vs_form_z_lit} illustrates that BB strengths $<1.0$ need recent star formation, and a dominant young stellar population, since the Balmer jump appears in the integrated spectrum only over a short period of time for CSF or declining SFHs. The BB strength between 1.0 and 2.0 can be explained by a variety of SFHs, and it is hard to constrain the formation redshift of those objects. This is evident in Fig.~\ref{fig_bb_vs_form_z_lit}, where the three SFHs considered here intersect at several points. On the other hand, BB $\ga 2.0$ cannot be explained by CSF at $z >2$ if we ignore the extinction. BB strengths in the range 2.0–2.75 require recent quenching, indicating a recent formation redshift for the dominant stellar population relative to the epoch of observation.

As shown in Figs.~\ref{fig_bb_vs_z_lit} and \ref{fig_bb_vs_form_z_lit}, a significant fraction of objects with BB $\ga 2.75$ appear at $z<5.0$. Such strong BB values can result from recent quenching, possibly after a burst. Alternatively, such strong BB strengths can also be achieved at $z<5.0$ through the passive evolution of an SSP formed at a very high redshift ($z \ga 10$). This suggests the possibility of forming the major fraction of stellar mass during an intense, bursty episode of star-formation in the early universe, followed by a rapid decline in the star formation rate, as demonstrated by \citet{Glazebrook+2024a, 2024MNRAS.534..325C} through detailed spectral fitting and reconstruction of the SFHs of QGs at $z=3-4$. The prevalence of BB candidates (BB$>1.5$) at $z\sim7.5-10$ further supports this scenario (see the IB tracks originating at formation redshifts $z=10$ and 12, which pass through the BB candidates at $z=7.5-10$ in Fig.~\ref{fig_bb_vs_form_z_lit}).

Independently of redshift, age, and formation redshift a maximum of BB $\sim 3$ can be reached with normal stellar populations in an IB without extinction (see Fig.~\ref{fig_bb_vs_form_z_lit}). Although strong BB values at $z\sim3$–4 can be explained with substantial extinction, this scenario is unlikely at $z>6$. There is a small number of sources with BB$>3.0$ present at $z>5$, including some with the strongest BB values. Most of these objects are indeed known LRDs, where the BB strength is explained by non-standard scenarios such as dense gas-rich envelopes (like BH$\star$) or SMSs \citep{2025ApJ...980L..27I, 2025MNRAS.tmp.1770J, 2025arXiv250316596N, 2025A&A...701A.168D, 2025arXiv250712618N,Chisholm2026Little-Red-Dots}. Such models could explain very strong BBs at any redshift, provided the timescale for the formation of these objects is short compared to cosmic time. For SMSs, this timescale can be of order of $\sim 1-2$ Myr or less \citep[see e.g.][]{2018MNRAS.478.2461G,Roman-Garza2026}. For this reason, a constant BB strength across redshift is shown in Figs.~\ref{fig_bb_vs_z_lit} and \ref{fig_bb_vs_form_z_lit}.

\subsection{Caveats --Reliability of BBs}
As in similar analyses, this study is subject to several caveats, including uncertainties in photometric redshifts, sample selection limitations, effects of photometric band properties, and the emergence of weak lines at young ages, which are discussed below.

\subsubsection{Reliability of photometric redshift}
As this study relies on photometry, a key limitation is the reliability of photometric redshift estimates. While recent studies find good agreement with spectroscopic redshifts \citep[e.g.,][]{Duan_2024a}, photometric redshifts may be  overestimated, especially at $z>8$ \citep{Fujimoto_2023a, Hainline_2024b, 2024ApJ...969L...2F}. However, the sample size and selection criteria outlined in Secs.~\ref{subsec_redshift_source_sel}, \ref{subsec_phot_spec_sample}, and Appendix~\ref{Appedix_z_sed_fits_spec} are sufficient to largely mitigate this issue. Nevertheless, some candidates in the highest redshift bin ($z=9.2$–10) may not be genuine BB sources; they could instead be emission-line galaxies at $z\sim8.5$, where line emission boosts the F444W flux and mimics a BB, or dusty galaxies at $z=3$–4 misclassified as Lyman-break galaxies at $z=9.2$–10. The limited number of filters longward of the break at z>9 makes it difficult to exclude such interlopers, allowing multiple solutions to coexist in SED fitting. Interloper contamination may bias the median BB in the highest redshift bin, such that the true median is likely somewhat lower than our estimate.

\subsubsection{Sample selection}
A relatively stringent detection threshold is applied, requiring $3\sigma$ detections in the BB-probing bands in all redshift bins, to ensure the reliability of the BB measurements. However, this strong detection threshold might lead to losing a notable fraction of objects that are intrinsically faint or affected by significant extinction. As BB shows no statistically significant strong correlation with absolute UV or V-band magnitude, intrinsically faint sources that may be missed could span the full range of BB strengths; therefore, their exclusion is unlikely to bias our results. If faint, potentially younger galaxies are preferentially missed, particularly in the highest-redshift bins, the median BB strength may be slightly overestimated. Conversely, if heavily obscured systems with strong BB features (e.g., dusty star-forming galaxies) fail to meet detection thresholds blueward of the break, the median BB may be underestimated. However, both effects are expected to be minor and unlikely to affect our conclusions.

\subsubsection{Photometric uncertainty and effects of bands}
Individual photometric uncertainties are not incorporated into the statistical analysis; instead, a median uncertainty for the sample is presented in the relevant figures. With $3\sigma$ detections in BB-probing bands for all candidates, neglecting individual uncertainties is unlikely to affect the results.

The sensitivity and bandwidth vary across the photometric bands. Among the six bands considered for BB measurements, F200W is the most sensitive (29.1 mag for a 10 ks exposure), followed by F150W (29.0 mag), F356W (28.7 mag), F277W (28.6 mag),  F444W (28.4 mag), and F410M (28.0 mag). Sensitivity differences between adjacent bands are small, and the band blueward of the break is generally more sensitive than the redder one for same exposure time, except for cases involving the F410M medium band. Overall, these sensitivity differences are expected to be negligible. Broadband filter bandwidths increase with wavelength; however, when combined with cosmological redshifting, appropriately chosen filters probe similar rest-frame wavelength intervals across the redshift ranges considered in this study.

In general, BB estimates derived from colors deviate from the D4000 Binggeli definition at older ages ($\gtrsim500$ Myr), due to the steep declining continuum and the inclusion of Mg II lines in the blue filter. This effect is illustrated in Fig.~\ref{fig_bb_diff_ind_met_inst_csf} and is metallicity dependent, with the discrepancy decreasing toward lower metallicity. Although these effects cannot be fully avoided, the highest redshift bins are subject to age constraints imposed by cosmology.

\subsubsection{Effect of other weak lines}
The redshift windows adopted in this analysis effectively avoid strong emission lines, which are typically observed in classical emission-line galaxies. However, at very young ages, various other lines including other Balmer series lines (like $\hg$, $\hd$, etc), \Neiii, \Hei, \Oiiill, and \Heii\ can be present \citep{2024ApJ...976..193R}. Even though these lines are not as strong as other optical emission lines like $H{\beta}$, their collective effect can boost the flux on the redder filter probing the break at younger ages. This is evident in Fig.~\ref{fig_bb_diff_ind_met_inst_csf}, where color-based BB estimates exceed the D4000 Binggeli definition. The strength of these lines increases with decreasing metallicity at fixed age. Therefore, in the highest redshift bins, where galaxies are expected to be younger and of lower metallicity, the median BB may be slightly overestimated due to these lines.

\section{Conclusions}
\label{sec_conclusion}

We investigated the observed and physical properties of a large sample of galaxies at $z=3.5$–10 using publicly available JWST photometric surveys, with particular emphasis on a systematic analysis of their Balmer break (BB) properties.To measure the BB strength and identify strong BB galaxies, we used adequate photometric filter combinations that can probe the break in redshift intervals, 3.5 - 4.0, 5.4-5.8, 7.1-8.0, 8.0-9.2, and 9.2-10, in a way that the filters are least affected by  emission lines. We divided our sample into two categories, the photometric and spectroscopic samples, where the redshift is determined from SED fitting for the former and measured from spectroscopy for the latter. We examined the general properties and trends of BB galaxies within the aforementioned redshift intervals and considered the broader implications of galaxy formation and evolution. Our main conclusions are the following: 

   \begin{enumerate}
      \item Photometric BB measurements in the selected redshift intervals are comparable to the spectroscopic indices D4000 Binggeli and D4000 Kuruvanthodi, and offer a larger dynamical range, thereby providing stronger constraints on stellar populations.
      \item We characterize the distribution of observed BB strengths over $z=3.5$–10, finding that it is well described by a distinct Gaussian distribution at each redshift bin. The average BB strength decreases with increasing redshift, likely due to the progressive increase in the mean stellar age from high to low redshift. The median break strength decreases from 0.5 mag to 0.13 mag (1.5 to 1.1 in flux ratio) between $z=3.5$ and 10, consistent with spectroscopic estimates \citep{2024ApJ...976..193R, Langeroodi+2024a}.  
      \item The break strength measured from the photometry is directly converted to age, assuming Constant Star Formation (CSF) and Instantaneous Burst (IB) Star Formation Histories (SFH). The age of the average stellar population decreases from 350 (50) Myr at $\sim z=3.5$ to 20 (10) Myr at $\sim z=10$ for the CSF (IB) scenario. In all redshift bins, we identify objects with BB strengths that cannot be reproduced by CSF models and instead require IB, more complex SFHs, or other explanations. This suggests that SFHs are diverse at high redshift and that a substantial fraction of galaxies are not in an actively star-forming phase at early epochs.
      \item We report moderate-to-strong positive correlations of BB strength with stellar mass, age, extinction, and UV slope ($\beta$) across all redshift bins, and corresponding negative correlations with $EW(\mathrm{H}\alpha)$ and $EW(\mathrm{H}\beta)$. However, a significant correlation between BB strength and sSFR is observed only in the lowest and highest redshift bins. No strong correlation is observed between BB strength and SFR, $M_{\mathrm{UV}}$, or $M_{V}$, suggesting that BB galaxies span all magnitude ranges.
      \item We compare our BB measurements with literature samples and find a consistent redshift distribution, with BB galaxies or candidates up to $z \sim 10$. The strongest BB sources occur at $z=3.5$–4 and $z\sim7$–10; the former include quiescent galaxies (with or without extinction) and LRDs, while the latter are predominantly LRDs. By comparing the BB evolution from the formation redshift for multiple SFHs with the observed candidates, we find that strong BB sources at $z=3.5$–4 can be explained either by recent quenching (at $z\sim5-6$) or by early formation ($z\geq10$) followed by passive evolution. Most strong BB candidates (BB$>3$) at $z<5$ are consistent with quenching combined with extinction. However, extremely strong breaks (BB>5) require exotic scenarios like SMS, AGN, or BH$\star$, and are likely LRDs.
   \end{enumerate}

\begin{acknowledgements}
      This work was supported by the Swiss National Science Foundation. A.K. acknowledges support from the Visvesvaraya PhD Scheme (Postdoctoral Fellowship) of the Ministry of Electronics and Information Technology (MeitY), Government of India. C.C. acknowledges support from the Swiss National Science Foundation through grant $200020\_192039$ (PI: Corinne Charbonnel). Some of the data products presented in this work were retrieved from the Dawn JWST Archive (DJA). DJA is an initiative of the Cosmic Dawn Center (DAWN), which is funded by the Danish National Research Foundation under Grant DNRF140. We thank Gabe Brammer for making this resource publicly available. We thank Prof. Pascal A. Oesch for useful discussions and suggestions.
\end{acknowledgements}

\bibliographystyle{aa}
\bibliography{aanda}

\begin{appendix}

\onecolumn
\section{SED fitting and mock simulations}
\label{append_cigale_para_space_sed_fits_sec}

\begin{table*}[ht!]
    \caption{The parameter space used in different modules on CIGALE to perform SED fits}
    \begin{tabular}{ c | c | c }
    \hline
    \hline
     & & Delayed $\tau$ model \\
    \hline 
            & $tau\_main$  & 2.0, 5.0, 14, 38, 100, 268, 713, 1900, 5063, 13489\\
            &              &  \\
            &              & 1, 2, 4, 6, 8, 12, 18, 26, 37, 54, 78,\\
            & $age\_main$  & 113, 164, 236, 342, 494, 713, 1030, 1487,\\ 
            &              & 2148, 3102, 4479, 6468, 9341, 13489\\
 sfhdelayed &              &  \\
            & $tau\_burst$ & 50.0 \\
            & $age\_burst$ & 1.0 \\
            & $f\_burst$   & 0 \\
            & $sfr\_0$ & 1.0 \\
            & normalise & True \\
    \hline
         & imf & 0 (Salpeter) \\
    bc03 & metallicity &  0.004\\
         & $separation\_age$ & 10 \\
    \hline
            & logU & -4.0, -3.0, -2.0, -1.0\\
            & zgas & 0.004\\
            & ne & 100 \\
    nebular & $f\_esc$ & 0.0 \\
            & $f\_dust$ & 0.0 \\
            & $lines\_width$ & 300.0 \\
            & emission & True \\
    \hline
                                   &                & 0, 0.1, 0.2, 0.3, 0.4, 0.5,\\
                                   & $E\_BV\_lines$ & 0.6, 0.7, 0.8, 0.9, 1.0, \\
                                   &                & 1.1, 1.2, 1.3, 1.4, 1.5 \\
                                   &                & \\
                                   & $E\_BV\_factor$ & 0.44 \\
                                   & $uv\_bump\_wavelength$ & 217.5\\
                                   & $uv\_bump\_width$ & 35.0 \\
    $dustatt\_modified\_starburst$ & $uv\_bump\_amplitude$ & 0.0 \\
                                   & $powerlaw\_slope$ & -0.5, -0.4, -0.3, -0.2, -0.1\\
                                   & $Ext\_law\_emission\_lines$ & 1 \\
                                   & Rv & 3.1 \\
                                   & filters & $B\_B90 \& V\_B90 \& FUV$ \\
    \hline
                            & $beta\_calz94$ & True \\
                            & D4000 & True \\
                            & IRX & True \\
    $restframe\_parameters$ & $EW\_lines$ & $372.7/1.0 \& 486.1/1.0$ \\
                            &          & $\& 500.7/1.0 \& 656.3/1.0$ \\
                            & $luminosity\_filters$ & $FUV \& V\_B90$ \\
                            & $colours\_filters$ & $FUV-NUV$ \\
    \hline
    redshifting & redshift & 0.0 -- 12 (with an increment of 0.05) \\
    \hline
    \end{tabular}
    \label{append_cigale_para_space_sed_fits}
\end{table*}

\FloatBarrier

\begin{table*}[ht!]
    \caption{The parameter space used in different modules on CIGALE to generate mock galaxy templates.}
    \begin{tabular}{ c | c | c | c }
    \hline
    \hline
     & & Constant Star Formation & Instantaneous Burst \\
    \hline 
            & $tau\_main$ & 99999999999999999999.99 & 0.0000001  \\
            & $tau\_burst$ & 0.001 & 0.001  \\
            & $f\_burst$ & 0.0 & 0.0  \\
            &  & 1, 2, 5, 10, 20, 50, 100, 200,  & 1, 2, 5, 10, 20, 50, 100, 200, \\
    sfh2exp & age & 400, 600, 800, 1000, 1200, 1400,  & 400, 600, 800, 1000, 1200, 1400,  \\
            &  & 1600, 1800, 2000, 2500, 3000, 3500, & 1600, 1800, 2000, 2500, 3000, 3500,  \\
            &  & 4000, 5000, 6000 & 4000, 5000, 6000 \\
            & $burst\_age$ & 1.0 & 1.0  \\
            & $sfr\_0$ & 1.0 & 1.0  \\
            & normalise & True & True \\
    \hline
         & imf & 0 (Salpeter) & 0 (Salpeter) \\
    bc03 & metallicity & 0.0001, 0.0004, 0.004, 0.008,  & 0.0001, 0.0004, 0.004, 0.008,  \\
         & & 0.02, 0.05 & 0.02, 0.05 \\
         & $separation\_age$ & 10 & 10 \\
    \hline
            & logU & -2.0 & -2.0 \\
            & zgas & 0.0004, 0.004, 0.008, 0.020, 0.051 & 0.0004, 0.004, 0.008, 0.020, 0.051 \\
            & ne & 100 & 100 \\
    nebular & $f\_esc$ & 0.0 & 0.0 \\
            & $f\_dust$ & 0.0 & 0.0 \\
            & $lines\_width$ & 300.0 & 300.0 \\
            & emission & True & True \\
    \hline
                                   & $E\_BV\_lines$ & 0.0 & 0.0 \\
                                   & $E\_BV\_factor$ & 1.0 & 1.0 \\
                                   & $uv\_bump\_wavelength$ & 217.5 & 217.5 \\
                                   & $uv\_bump\_width$ & 35.0 & 35.0 \\
    $dustatt\_modified\_starburst$ & $uv\_bump\_amplitude$ & 0.0 & 0.0 \\
                                   & $powerlaw\_slope$ & -0.5 & -0.5 \\
                                   & $Ext\_law\_emission\_lines$ & 3 & 3 \\
                                   & Rv & 2.93 & 2.93 \\
                                   & filters & $B\_B90 \& V\_B90 \& FUV$ & $B\_B90 \& V\_B90 \& FUV$ \\
    \hline
                            & $beta\_calz94$ & True & True \\
                            & D4000 & True & True \\
                            & IRX & True & True \\
    $restframe\_parameters$ & $EW\_lines$ & $372.7/1.0 \& 486.1/1.0$ & $372.7/1.0 \& 486.1/1.0$ \\
                            &          & $\& 500.7/1.0 \& 656.3/1.0$ & $\& 500.7/1.0 \& 656.3/1.0$ \\
                            & $luminosity\_filters$ & $FUV \& V\_B90$ & $FUV \& V\_B90$ \\
                            & $colours\_filters$ & $FUV-NUV$ & $NUV-r\_prime$ \\
    \hline
    redshifting & redshift & 1.0 -- 10 (with an increment of 0.1) & 1.0 -- 10 (with an increment of 0.1) \\
    \hline
    \end{tabular}
    \label{append_cigale_para_space}
\end{table*}

\FloatBarrier 
\section{Photometric sample selection}
\label{Appedix_z_sed_fits_spec}
\begin{figure*}[ht!]
    \centering
    \includegraphics[width=1\linewidth]{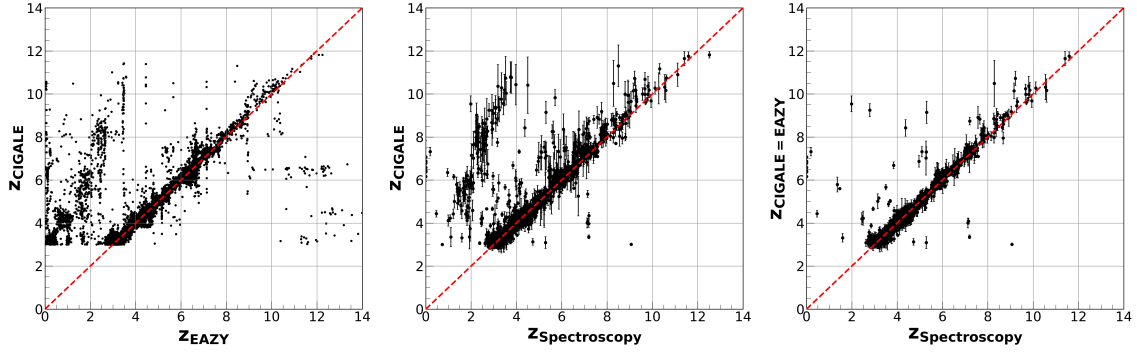}
    \caption[Comparison of redshift estimation from different SED fitting codes and spectroscopy]{Comparison of redshift estimation from different SED fitting codes and spectroscopy. The one-to-one line is shown in red (dashed line). \textbf{Left:} Photometric redshift estimation from CIGALE versus EAZY. \textbf{Middle:} Photometric redshift estimation from CIGALE versus spectroscopic redshift estimations. \textbf{Right:} Photometric redshift estimations from CIGALE, which agree with EAZY estimations within median redshift uncertainties in the sample versus spectroscopic estimations.}
    \label{fig_z_cigale_eazy_spec}
\end{figure*}

To understand the reliability of redshift estimations, we first compare the CIGALE redshift estimations from this work with EAZY redshift estimations from \citep{2024MNRAS.533.1808W}. Fig. \ref{fig_z_cigale_eazy_spec} shows the comparison of redshift estimations from CIGALE and EAZY. In general, redshift estimations from CIGALE and EAZY show an agreement for the majority of the sample. But there are some outliers also, where sometimes CIGALE gives a high redshift solution while EAZY does not, and vice versa. These outliers are unavoidable in large photometric studies like the one presented here.

To better understand the performance of CIGALE to estimate the redshift, we compared the redshift estimations from CIGALE to the spectroscopic redshift measurements for a sample of objects for which spectra are available, and this is also shown in Fig. \ref{fig_z_cigale_eazy_spec} (middle). For the majority of the sample, photometric redshift estimations from CIGALE are in agreement with the spectroscopic redshift estimations. While CIGALE overestimates the redshift in some cases. To further reduce the interlopers in the photometric sample, we checked whether the sources for which the redshift estimations from the two different SED fitting codes are similar show a better agreement with spectroscopy or not. From the full spectroscopic sample, we made a sub-sample of objects in which the difference between CIGALE and EAZY estimations is less than or equal to 0.2, where 0.2 corresponds to $\sim1.2$ times the quadrature-combined median uncertainty of the two photometric redshift estimates for the entire photometric sample at $z\geq3$. This is also shown in Fig. \ref{fig_z_cigale_eazy_spec} (Right), and it is evident that this sample shows fewer interlopers than the previous ones. We notice that in some cases, EAZY fits the SEDs with a template dominated by emission lines, while CIGALE fits using a template with a Balmer break. Comparing the redshift estimations from two codes avoids this degeneracy for many of the sources and reduces the interlopers in the sample. Apart from that, \citet{2024MNRAS.533.1808W} indicated that redshift estimations from EAZY with a few templates are slightly better than using the SED fitting code, which relies on the complete parameter space.\\ 

\section{Mock galaxy simulation results}

\begin{figure*}[ht!]
    \centering
    \includegraphics[width=1\linewidth]{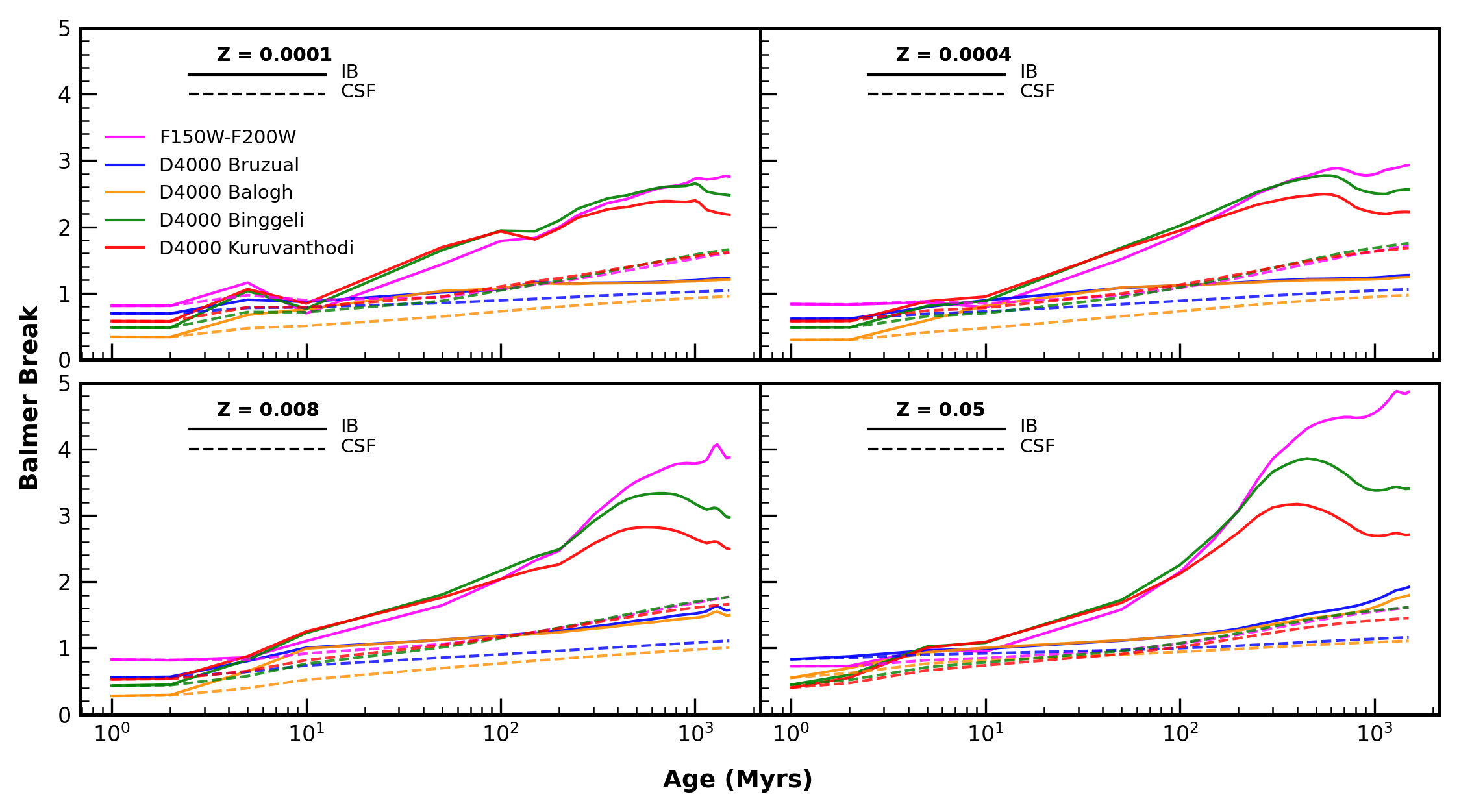}
    \caption[Predicted BB as a function of age for different metallicity ranges for the Instantaneous burst scenario]{BB estimation for different metallicity ranges for both the IB (solid lines) and CSF (dashed lines) scenario. The blue color indicates the D4000 break definition from \citet{1983ApJ...273..105B}, the orange color indicates the definition from \citet{1999ApJ...527...54B}, the green color indicates the definition from \citet{2019MNRAS.489.3827B}, the red color indicates the definition used in \citet{2023A&A...674A.140K}, and the magenta color indicates the estimations from F150W-F200W color.}
    \label{fig_bb_diff_ind_met_inst_csf_p2_appendix}
\end{figure*}


\begin{table}[ht!]
    \centering
    \caption{The slope and intercept estimated to measure the age of the dominant stellar populations from the BB strength for the IB and CSF SFH.}
    \begin{tabular}{l | lll | lll}
    \hline
    redshift &           &      IB       &             &           &     CSF       &   \\ 
             & Slope (m) & Intercept (c) & BB range    & Slope (m) & Intercept (c) & BB range \\
    \hline
                   & 4.016& 0.2658& BB<0.19            &      9.307&         0.1759& BB<0.1\\
        $3.5-4.0$  & 1.379& 1.0480& $0.19\leq$ BB <1.4 &      3.130&        1.1850& $0.1\leq$ BB <0.61\\
                   & -3.153&7.5820& BB$\geq1.4$        &      1.222&         2.3560& $0.61\leq$ BB $\leq0.74$\\
                   &      &       &                    &           &               & \\
        $5.4-5.8$  & 4.197& 0.3760& BB<0.19            &     10.580&         0.3830& BB<0.07\\
                   & 1.143& 1.1650& BB$\geq0.19$       &      2.846&         1.3230& $0.07\leq$ BB $\leq0.65$\\
                   &      &       &                    &           &               & \\
        $7.1-8.0$  & 3.579& 0.3639& BB<0.19            &      7.559&         0.4795& BB<0.09\\
                   & 1.385& 1.0650& BB$\geq0.19$       &      3.392&         1.1670& $0.09\leq$ BB $\leq0.52$\\
                   &      &       &                    &           &               & \\
        $8.0-9.2$  & 2.859& 0.6536& BB<0.19            &      5.773&         1.0760& BB<0.09\\
                   & 0.9464&1.2050& BB$\geq0.19$       &      2.465&         1.4370& $0.09\leq$ BB $\leq0.42$\\
                   &      &       &                    &           &               & \\
        $9.2-10.0$ & 3.292& 0.3786& BB<0.19            &      6.674&         0.5481& BB<0.09\\
                   & 1.585& 1.0020& BB$\geq0.19$       &      3.925&         1.0700& $0.09\leq$ BB $\leq0.42$\\
    \end{tabular}
    \label{tab_BB_age}
\end{table}


\begin{figure*}[ht!]
    \centering
    \includegraphics[scale = 0.35]{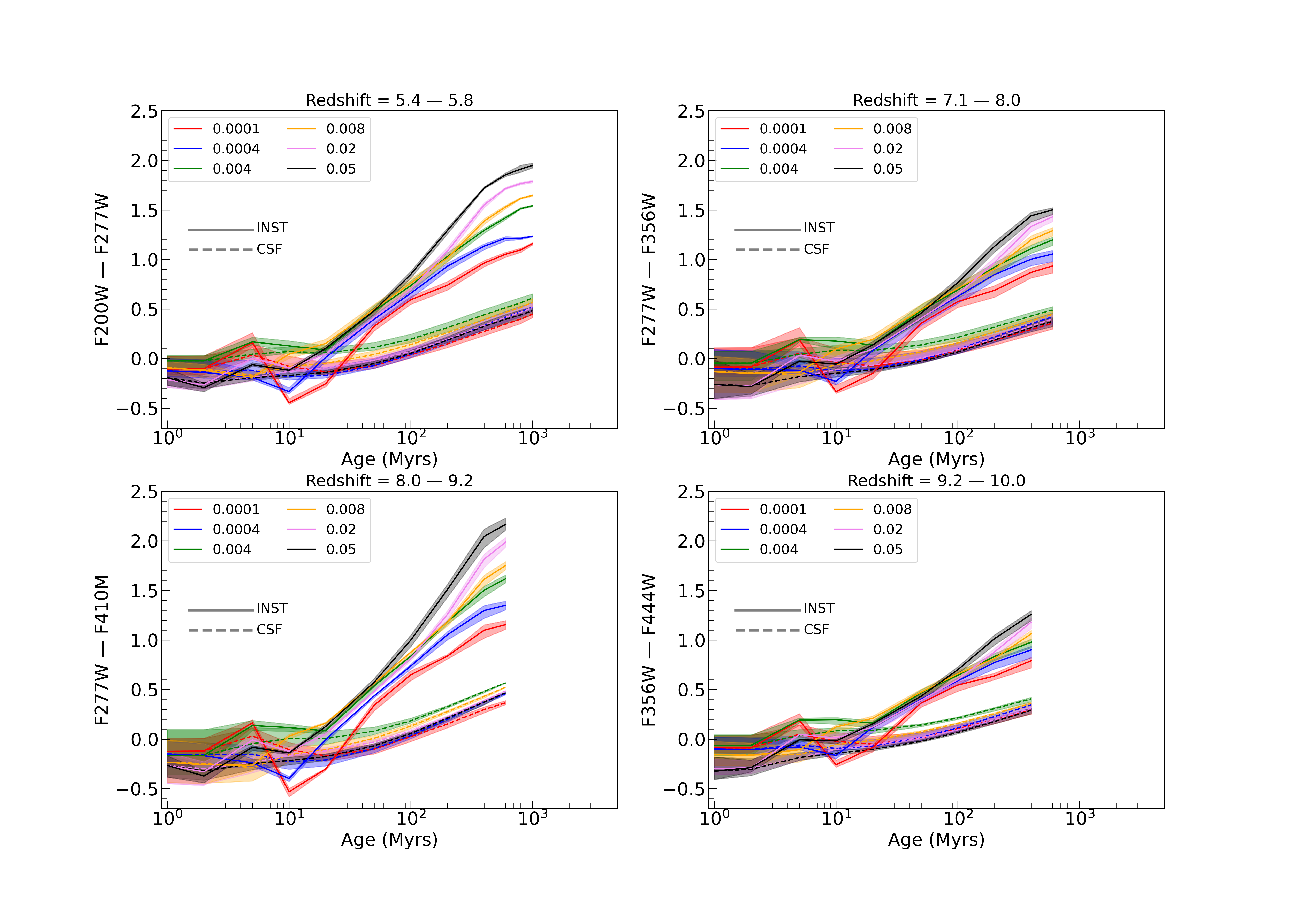}
    \caption{BB probing color versus age for different metallicity ranges and for CSF and IB scenario. The shaded region around each metallicity indicates the spread in BB due to the redshift change within the corresponding redshift bin. The solid line indicates the IB scenario and the dashed line  the CSF scenario.}
    \label{fig_bb_color_met_inst_csf_z5p4_10p0}
\end{figure*}



\clearpage
\newpage
\FloatBarrier
\section{Observed and derived properties}

\begin{figure}[ht!]
    \centering
    \includegraphics[scale=0.5]{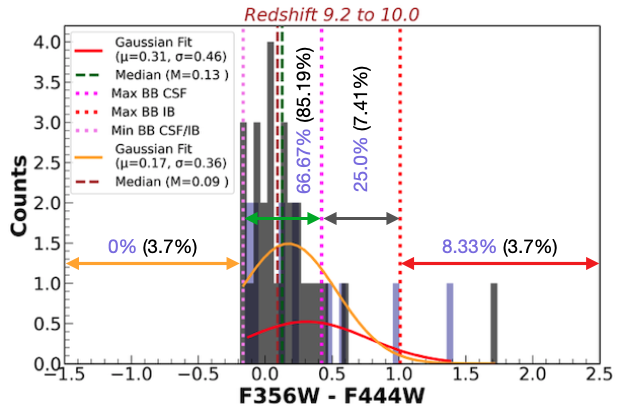}
    \caption{BB distribution of objects in $z=9.2-10$ redshift bin. A Gaussian fit performed for the distribution is shown in the red and yellow lines for photometric and spectroscopic samples, respectively. The median value of the distribution is shown in green and brown dashed lines for photometric and spectroscopic samples, respectively. The maximum BB expected from the CSF and IB (without extinction, E(B-V) = 0) are shown in magenta and red dotted lines, respectively. The minimum BB expected from CSF/IB is shown in a violet dashed line. The fractions of objects with BB below the minimum model prediction, consistent with the CSF scenario, consistent with only IB scenario, and inconsistent with classical models are reported for the photometric (spectroscopic) sample.
    }
    \label{fig_hist_gauss_count_part2}
\end{figure}


\begin{figure*}[ht!]
    \centering
    \includegraphics[width=1\linewidth]{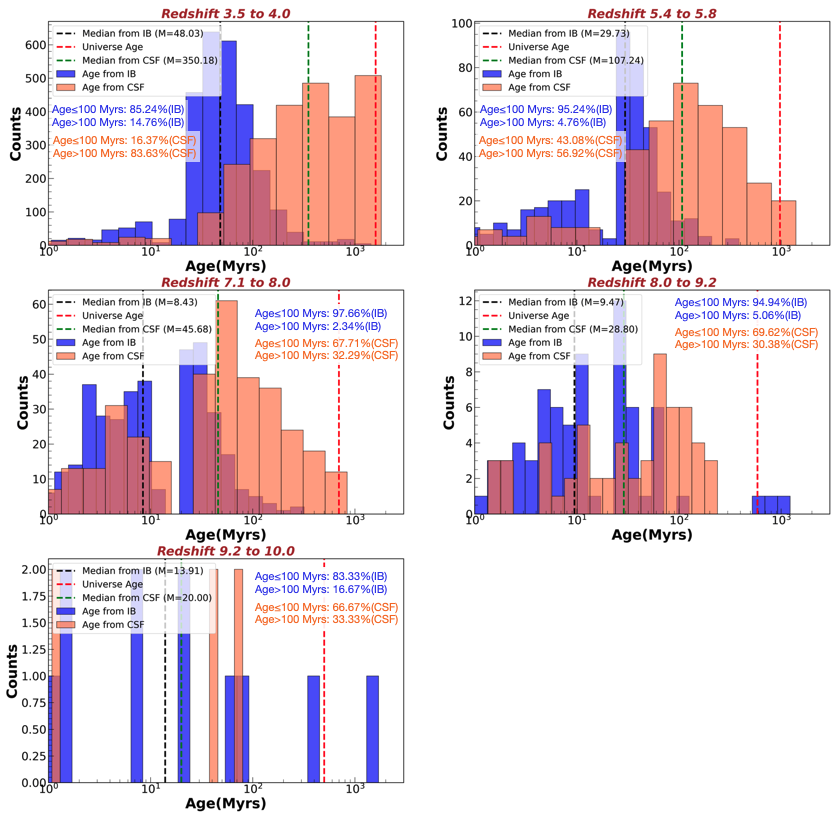}
    \caption{Age distribution of objects in different redshift bins using IB (Blue) and CSF (Brown) scenario for the photometric sample. The median age is shown in the green and black vertical lines for IB and CSF, respectively. The maximum allowed age of the Universe by cosmology in each redshift bin is shown in the red dashed line. In some cases, the CSF scenario provides unrealistic (very high) ages for the strong BB candidates in the highest redshift bins. Those sources are not included in the histogram.
    }
    \label{fig_hist_age_inst_csf}
\end{figure*}


\begin{figure*}[ht!]
    \centering
    \includegraphics[width=1\linewidth]{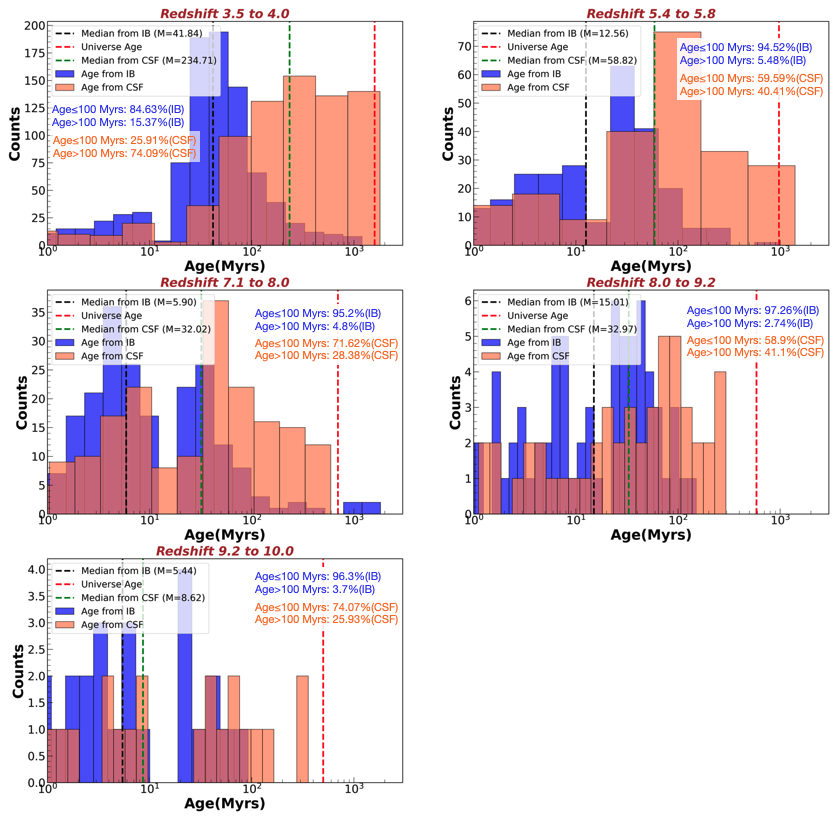}
    \caption{Age distribution of objects in different redshift bins using IB (Blue) and CSF (Brown) scenario for the spectroscopic sample. The median age is shown in the green and black vertical lines for IB and CSF, respectively. The maximum allowed age of the Universe by cosmology in each redshift bin is shown in the red dashed line. In some cases, the CSF scenario provides unrealistic (very high) ages for the strong BB candidates in the highest redshift bins. Those sources are not included in the histogram.
    }
    \label{fig_hist_age_inst_csf_spec}
\end{figure*}



\begin{figure*}[ht!]
    \centering
    \includegraphics[width=1\linewidth]{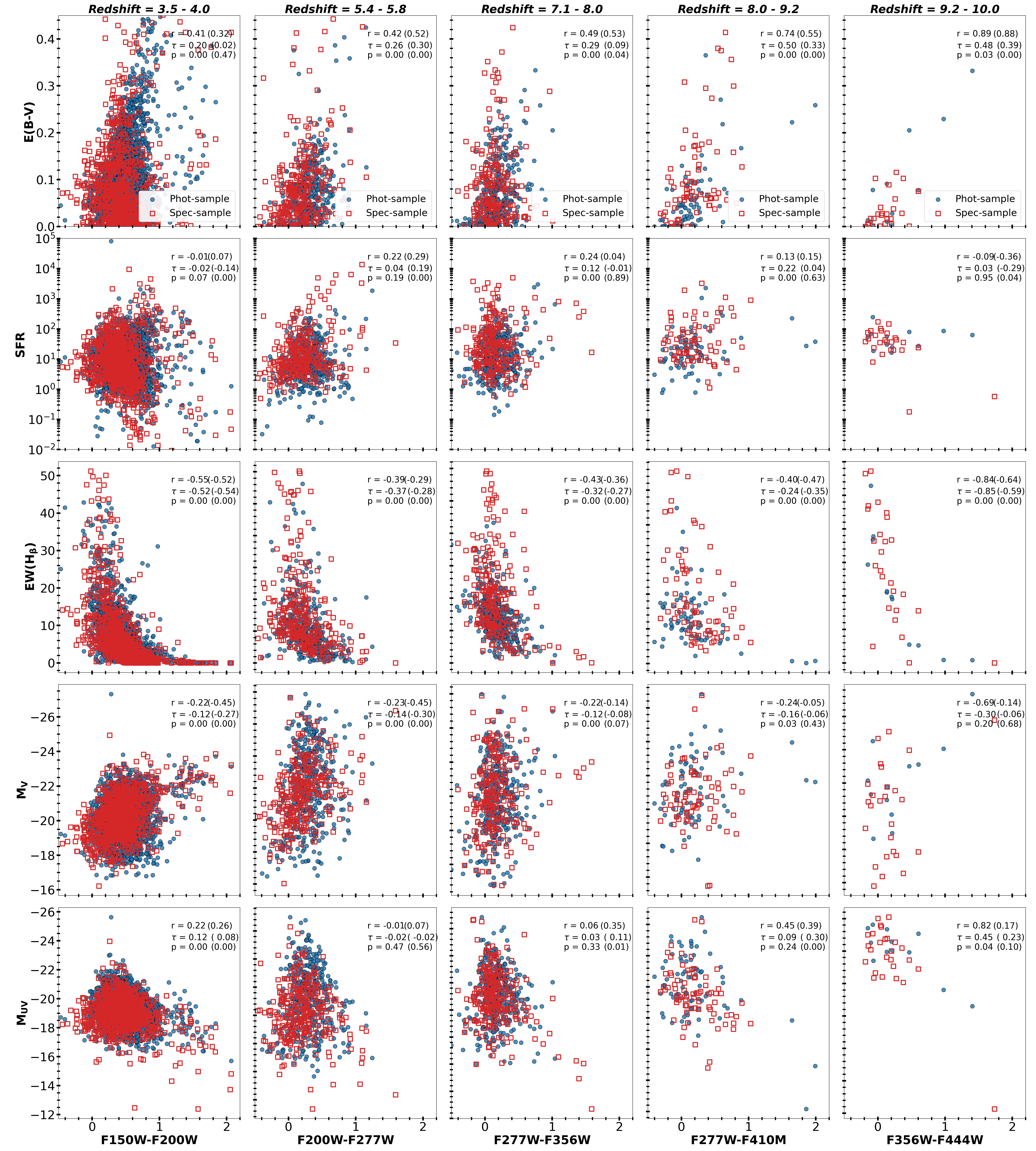}
    \caption{Correlation of Balmer break strength with various physical parameters for the photometric and spectroscopic sample (PART-2). The blue dot represents the photometric sample, while the red squares represent the spectroscopic sample. Redshift increases from left to right (in the order of 3.5-4, 5.4-5.8, 7.1-8.0, 8.0-9.2, and 9.2-10). Each row corresponds to various physical parameters. From top to bottom, it indicates E(B-V), SFR, EW$(H_{\alpha})$, $M_{V}$, and $M_{UV}$. Within each plot, Pearson and Kendall correlation coefficients are given for photometric (spectroscopic) sample.
    }
    \label{fig_param_vs_bb_pear_kend_p2}
\end{figure*}


\end{appendix}

%
%

\end{document}